\documentclass[final,hidelinks,onefignum,onetabnum]{siamart251216}
\usepackage[T1]{fontenc}
\usepackage{lmodern,microtype}
\usepackage{amsfonts,amssymb,bm,booktabs,array,graphicx,algpseudocode,placeins}
\usepackage{needspace}
\graphicspath{{./}{figures/}}
\DeclareGraphicsExtensions{.pdf,.png,.jpg}
\newsiamremark{remark}{Remark}

\newcommand{\dt}{\Delta t}
\newcommand{\fvec}{\bm{f}}
\newcommand{\uvec}{\bm{u}}
\newcommand{\kvec}{\bm{k}}
\newcommand{\cs}{c_s}
\renewcommand{\Re}{\mathrm{Re}}
\newcommand{\Ma}{\mathrm{Ma}}

\newcommand{\Swall}{S_{\mathrm{wall}}}
\newcommand{\Smarch}{S_{\mathrm{march}}}
\newcommand{\Swork}{S_{\mathrm{work}}}
\newcommand{\Stotal}{S_{\mathrm{total}}}
\headers{ACCELERATING DUAL-TIME STEPPING}{Y. DING, Y. YU, Y. YU, AND H. ZHANG}
\title{Residual correction for accelerating dual-time-stepping lattice Boltzmann methods\thanks{{\bfseries Funding:} This work was supported by the National Natural Science Foundation of China (grant 12101527), the Natural Science Foundation of Hunan Province (grant 2026JJ60005), the Science and Technology Innovation Program of Hunan Province (grant 2026RC3172), and the 111 Project (grant D23017).}}
\author{Yanhui Ding\orcid{0009-0008-0420-6854}\thanks{School of Mathematics and Computational Science, Xiangtan University, Xiangtan 411105, China.}
\and Yaolong Yu\orcid{0009-0002-9599-4353}\footnotemark[2]
\and Yuan Yu\orcid{0000-0001-5125-2492}\thanks{Corresponding author. School of Mathematics and Computational Science, Xiangtan University; National Center for Applied Mathematics in Hunan; Hunan Key Laboratory for Computation and Simulation in Science and Engineering, Xiangtan 411105, China (\email{yuyuan@xtu.edu.cn}).}
\and Hao Zhang\orcid{0009-0005-5307-1229}\footnotemark[2]}
\hypersetup{pdftitle={Residual correction for accelerating dual-time-stepping lattice Boltzmann methods},pdfauthor={Yanhui Ding, Yaolong Yu, Yuan Yu, Hao Zhang}}
\begin{document}
\maketitle
\begin{abstract}
For lattice Boltzmann simulations of unsteady flows at low Reynolds numbers,
dual-time stepping decouples the physical time step from lattice streaming, yet
conserved-moment iteration errors decay slowly at large steps. We develop a
residual correction method based on the conserved-moment structure of the
collision operator. With fixed step histories and both relaxation rates equal to
unity, post-collision populations depend on incoming populations only through
reconstructed conserved moments, so one collision eliminates local perturbations
with zero conserved moments. Periodic rest-state linearization shows that, at
large physical steps, the contraction factor of the uniform conserved mode
approaches unity, while its moment residual scales inversely with the step
relative to macroscopic error. On periodic domains, combining this scaling with
linearized acoustic propagation and viscous diffusion yields a macroscopic
auxiliary system driven by the conserved-moment residual and solved in Fourier
space for low-wavenumber errors. For solid walls, reference-state linearization
of the complete sweep, including executed bounce-back updates, yields a sparse
kinetic correction. Both auxiliary systems compute increments from the full
nonlinear residual, and we prove fixed-point equivalence with the original
discrete system, preserving unsteady and nonlinear convective effects. Refining
the time step on fixed grids verifies second-order temporal self-convergence of
slowly varying periodic velocity fields in two and three dimensions. Under fixed
physical steps and tolerances, mean inner sweeps per step drop from hundreds in
three-dimensional periodic flow and thousands in a two-dimensional cavity to a
few.
\end{abstract}

\section{Introduction}\label{sec:intro}
Low-Reynolds-number flows can exhibit rich unsteady dynamics driven by internal
stresses, microstructure, or time-varying forcing: elastic stresses trigger
elastic turbulence in polymer solutions~\cite{GroismanSteinberg2000}, active
stresses sustain evolving vortices and defects in active
nematics~\cite{Thampi2014ActiveTurbulence}, and Newtonian flows with viscous
diffusion time $t_\nu=L^2/\nu$ comparable to the driving period $T$
($t_\nu/T=O(1)$) require time-accurate resolution even at
$\Re\ll 1$~\cite{Bruot2021}. On-grid lattice Boltzmann methods (LBMs) evolve $Q$
discrete-velocity populations through local collision, interpolation-free link
streaming, and local wall closures~\cite{LallemandEtAl2021,Yu2026}. Explicit
stepping, however, ties physical time increments to lattice streaming: under
nearly incompressible diffusive scaling
$\delta t_{\mathrm{ph}}\propto\delta x_{\mathrm{ph}}^2$, advancing across a
fixed fraction of $t_\nu$ on an $N^d$ mesh requires $O(N^2)$ steps and
$O(N^{d+2})$ site updates~\cite{LallemandEtAl2021}. Because relaxation rates and
lattice sound speed simultaneously govern viscous-stress recovery, lattice Mach
number $\Ma$, and near-wall accuracy, enlarging explicit steps requires
continuity and pressure modifications~\cite{Reis2020StressLBM,Gsell2021,Horstmann2022}.
The computational goal is therefore to decouple the physical time step from
lattice streaming while preserving local collision, exact link shifting, and
temporal accuracy at the chosen collision operating point.

Off-lattice implicit LBMs uncouple the velocity lattice from the spatial mesh,
discretize advection by finite differences, and solve each step via
Newton--Krylov--Schwarz iterations~\cite{Huang2015ImplicitLBM}. While lifting
the acoustic CFL limit, upwind stencils and ghost-node extrapolation introduce
spatial dissipation and sacrifice exact link shifts and local bounce-back;
solving the $Q N_\Omega$-unknown nonlinear system also yielded only a
$1.58$-fold wall-clock speedup despite a $6375$-fold step reduction in transient
Couette flow. On-grid dual-time stepping instead separates outer implicit
physical derivatives from matrix-free pseudo-time sweeps that preserve unit link
shifts and local wall closures~\cite{Gsell2020,ArmstrongPeng2021,JiLuo2026}.
Introducing the physical time derivative as a macroscopic source---with BDF1
startup and BDF2 thereafter rather than $Q$ population
derivatives~\cite{JiLuo2026}---keeps unit link shifts, lattice $\Ma$, and TRT
relaxation rates $\omega^+=\omega^-=1$ independent of $\dt$, resolves the linear
BDF source pointwise via trapezoidal half-source reconstruction without local
sub-iterations~\cite{GruszczynskiEtAl2023}, and confines temporal coupling to
$d+1$ conserved moments. Existing LBM formulations based on dual-time stepping
nevertheless report only four- to fivefold speedups in unsteady bluff-body
flows~\cite{Gsell2020,JiLuo2026}, because inner sweep counts grow with $\dt$ and
offset outer-step savings.

Inner convergence at large $\dt$ can be accelerated in population space or
through reduced moment equations. Full-population multigrid~\cite{Gsell2020}
transfers $Q$-component fields across grids, whereas Anderson
acceleration~\cite{WalkerNi2011,DeSterck2021AA} forms updates from a finite
residual history without exploiting collision rank structure. Macroscopic--kinetic
couplings instead propagate slow moment corrections through reduced systems:
asymptotic-preserving and micro--macro Schur methods eliminate microscopic
unknowns locally into diffusive or elliptic
equations~\cite{Peng2021Schur,Zhang2023KineticFD,HauckLaiuSchnake2025}, implicit
UGKS drives macroscopic predictions from kinetic
residuals~\cite{ZhuZhongXu2019,Xu2022UGKS}, and GSIS couples higher-order moment
closures into macroscopic equations~\cite{Su2020GSIS,ZengSuWu2023} with spectral
analysis guiding slow-mode damping~\cite{Dong2024Moments}. Yet fixed-point
moment consistency and rapid large-step contraction impose distinct algebraic
conditions~\cite{HauckLaiuSchnake2025}. Our low-Mach BDF sweep needs no rarefied
stress closures, but continuum macroscopic equations cannot directly replace its
discrete derivative: half-source reconstruction modifies local collision
derivatives and residual scaling~\cite{GruszczynskiEtAl2023}, and solid-wall
bounce-back acts inside each sweep before residual evaluation, so continuum
macroscopic wall conditions introduce boundary mismatch that degrades
convergence~\cite{Reis2020StressLBM,Dong2024Moments,LiuZhangZengWu2024}.

Analyzing the complete BDF sweep with step histories fixed explains why local
nonconserved-mode elimination does not prevent slow large-step convergence. At
unit TRT relaxation $\omega^+=\omega^-=1$, post-collision populations depend on
incoming populations solely through $d+1$ reconstructed conserved moments, so
the local collision derivative has rank at most $d+1$ and annihilates all local
zero-moment perturbations in one collision, while half-source reconstruction and
streaming still couple conserved moments spatially. Periodic rest-state
linearization reveals two coupled large-step effects: as $\dt$ grows, the
contraction factor of the uniform conserved mode approaches unity while its
projected residual shrinks as $O(\dt^{-1})$ relative to the reconstructed
macroscopic error, requiring both targeted slow-mode acceleration and
step-appropriate stopping tolerances.

Linearizing the complete-sweep residual about a reference state yields a unified
defect framework whose rank-$(d+1)$ collision factorization gives an exact Schur
relation between population and moment increments. On periodic domains, we
approximate the conserved-moment response in raw-moment space to form a
macroscopic shadow correction (MSC), solved in Fourier space on $d+1$ scalar
fields (three in 2D and four in 3D) and lifted to populations; we refer to this
periodic solver as the macroscopic-shadow-corrected lattice Boltzmann method
(MSC-LBM). At large $\dt$, MSC-LBM reduces the uniform-mode contraction factor
to $O(\dt^{-1})$ while axial Nyquist modes on even grids retain their baseline
magnitude. On wall-bounded domains, we differentiate the complete
sweep---including its ordered bounce-back updates---about a reference state to
form a sparse kinetic defect matrix $I-L$ whose LU factors are reused across
sweeps and steps, with residual-triggered relinearization. Under fixed history
and boundary data, positive step and transport parameters, and nonsingular
$I-L$, both branches preserve the discrete BDF fixed-point set.

The numerical experiments verify discrete trajectory consistency and
second-order temporal self-convergence of slowly evolving velocity fields on
periodic domains in two and three dimensions. At fixed physical time step $\dt$
and residual tolerance $\varepsilon=10^{-10}$, mean inner sweeps per step drop
from over $100$ and $2389$ (with initial Picard steps hitting the iteration cap)
to $3.2$ in $128^3$ periodic flow ($\Re=1$) and $5.8$ in a $64^2$ lid-driven
cavity ($\Re=10$); separate timing comparisons (\cref{sec:protocol}) measure
both physical marching time and total solver runtime including initialization
against explicit stepping.
\Cref{sec:framework,sec:dc} formulate the discrete BDF sweep and its large-step
bottleneck, \cref{sec:correctors} constructs MSC-LBM and the wall-bounded
reference kinetic correction, \cref{sec:results} presents the numerical
experiments, and \cref{sec:conclusions} summarizes the conclusions; extended
protocols and multigrid details are given in the supplementary materials.

\section{The discrete physical-step problem and baseline iteration}\label{sec:framework}
Each physical time step requires a solution of a discrete BDF lattice
problem with accepted histories held fixed. The physical index $n$ labels
accepted time levels, while the inner index $\ell$ counts sweeps toward that
step's fixed point. A unit link shift is part of an inner sweep; the physical
time advances only after the step satisfies the residual criterion in
\cref{sec:acceptance}.

\subsection{BDF sources and macroscopic reconstruction}\label{sec:operator}
In lattice units with unit spacing and unit acoustic streaming increment, the
physical step $\dt$ is chosen independently of the mesh. At unit reference
density, define the macroscopic state
$\bm m=(\hat p,\uvec)^{\mathsf T}$, where $\hat p=p/c_s^2$. The associated
low-Mach macroscopic description is given by~\cite{HeLuo1997,Dellar2003}
\begin{equation}\label{eq:macro-pde-pair}
 \partial_t\hat p+\nabla\!\cdot\uvec=0,\qquad
 \partial_t\uvec+(\uvec\!\cdot\!\nabla)\uvec
 =-c_s^2\nabla\hat p+\nu\nabla^2\uvec.
\end{equation}
In \eqref{eq:macro-pde-pair}, $c_s$ is the lattice sound speed and $\nu$ the
kinematic viscosity; $\partial_t\hat p$ is higher order on the slow
incompressible trajectory yet remains part of the discrete evolution.

The time integration starts with one BDF1 step and advances subsequent physical
steps with BDF2~\cite{HairerWanner1996}. Approximating the negative physical-time
derivative $-\partial_t\bm m$ by BDF gives the effective source
$\bm s=\bm h_q-a_q\bm m$ of order
$q\in\{1,2\}$ with
\begin{equation}\label{eq:bdf2}
 (a_q,\bm h_q)=
 \begin{cases}
 (1/\dt,\bm m^n/\dt),&q=1,\\
 (3/(2\dt),(4\bm m^n-\bm m^{n-1})/(2\dt)),&q=2.
 \end{cases}
\end{equation}
The scalar $a_q>0$ weights the unknown state $\bm m$, while $\bm h_q$ stores
accepted histories $\bm m^n,\bm m^{n-1}$. Because $a=a_q$, $\bm h=\bm h_q$, and
boundary data $\Gamma$ stay fixed within one physical step, we drop the
new-time superscript on step unknowns.

At each site, populations $\fvec=(f_0,\ldots,f_{Q-1})^{\mathsf T}$ live on D2Q9
($d=2,Q=9$) or D3Q19 ($d=3,Q=19$) with $c_s^2=1/3$~\cite{Qian1992,ChenDoolen1998},
whose velocities $\bm c_k$ comprise the rest vector, coordinate links, and face
diagonals with weights $w_k$ equal to $(4/9,1/9,1/36)$ in 2D and
$(1/3,1/18,1/36)$ in 3D. For $\alpha\in\{1,\ldots,d\}$, raw-moment projection
$P$ and rest-state lift $E$ are defined by
\begin{equation}\label{eq:linear-lift}
 P_{0k}=1,\quad P_{\alpha k}=c_{k\alpha},\qquad
 E_{k0}=w_k,\quad E_{k\alpha}=w_kc_{k\alpha}/c_s^2,
 \qquad PE=I_{d+1}.
\end{equation}
With $I_{d+1}$ denoting the $(d+1)\times(d+1)$ identity matrix, second-order quadrature gives
$PE=I_{d+1}$ (\cref{supp:app:lattices}). Substituting $\bm s=\bm h-a\bm m$ into
the half-source rule $\bm m=P\fvec+\bm s/2$~\cite{Guo2002,Gsell2020} yields
\begin{equation}\label{eq:macro-reconstruct}
 d_a=1+a/2,\qquad
 \bm m=\mathcal R_{q,\bm h}(\fvec)=\frac{P\fvec+\bm h/2}{d_a},
 \qquad \bm s=\bm h-a\bm m.
\end{equation}
With the denominator $d_a$ in the reconstruction, the affine map
$\mathcal R_{q,\bm h}$
expresses both state and source through incoming raw moments.

\subsection{The complete discrete sweep}\label{sec:scheme}
Reconstructed moments $\bm m$ and sources $\bm s=(s_0,\bm s_u)^{\mathsf T}$
specify the incompressible equilibrium and source distributions,
\begin{align}
 f_k^{\rm eq}(\bm m)&=w_k\left[\hat p+
 \frac{\bm c_k\!\cdot\uvec}{c_s^2}+
 \frac{(\bm c_k\!\cdot\uvec)^2}{2c_s^4}-
 \frac{\uvec\!\cdot\uvec}{2c_s^2}\right],\label{eq:feq}\\
 \Phi_k(\bm m,\bm s)&=w_ks_0+w_k
 \left[\frac{\bm c_k-\uvec}{c_s^2}
 +\frac{(\bm c_k\!\cdot\uvec)\bm c_k}{c_s^4}\right]
 \!\cdot\bm s_u.\label{eq:source-distribution}
\end{align}
With pressure and momentum sources $s_0$ and $\bm s_u$, fourth-order quadrature
ensures $P\fvec^{\rm eq}=\bm m$ and $P\bm\Phi=\bm s$. In TRT
collision, even and odd parts $g_k^\pm=(g_k\pm g_{\bar k})/2$ with
$\bm c_{\bar k}=-\bm c_k$ relax at rates $\omega^\pm=1/\tau^\pm$ linked to
$\nu$ and magic parameter $\Lambda$ via~\cite{Ginzburg2008}
\begin{equation}\label{eq:trt}
 \nu=c_s^2(\tau^+-1/2),\qquad
 \Lambda=(\tau^+-1/2)(\tau^--1/2).
\end{equation}
The relaxation time $\tau^+$ sets the viscosity; for fixed $\tau^+$,
$\tau^-$ determines $\Lambda$. The post-collision populations are
\begin{equation}\label{eq:collision}
 \mathcal C_k=f_k-\omega^+(f_k^+-f_k^{{\rm eq},+})
 -\omega^-(f_k^--f_k^{{\rm eq},-})
 +(1-\omega^+/2)\Phi_k^++(1-\omega^-/2)\Phi_k^-.
\end{equation}
At our primary operating point $\omega^+=\omega^-=1$ ($\nu=1/6$,
$\Lambda=1/4$), $f_k=f_k^++f_k^-$ cancels incoming non-equilibrium terms in
\eqref{eq:collision}, leaving
\begin{equation}\label{eq:collision-operating}
 \mathcal C_k(\fvec)=f_k^{\rm eq}(\bm m)+\tfrac12\Phi_k(\bm m,\bm h-a\bm m),
 \qquad P\mathcal C=\bm m+\bm s/2=P\fvec+\bm s.
\end{equation}
Under \eqref{eq:collision-operating}, $\mathcal C_k(\fvec)$ depends on $\fvec$
solely through $\bm m$. The half-source contribution in macroscopic
reconstruction and the half-source contribution in collision combine to give
$P\mathcal C=P\fvec+\bm s$. Composing collision with push streaming
$(\mathcal Sg)_k(\bm x+\bm c_k)=g_k(\bm x)$ and affine boundary closure
$\mathcal B_\Gamma(\cdot)+\bm b_\Gamma$, with linear part $\mathcal B_\Gamma$
and offset $\bm b_\Gamma$, defines the complete sweep
\begin{equation}\label{eq:sweep}
 T(\fvec)=\mathcal B_\Gamma\mathcal S\mathcal C(\fvec)+\bm b_\Gamma,
 \qquad T(\fvec^\star)=\fvec^\star.
\end{equation}
Solutions $\fvec^\star$ define the physical-step fixed points; on periodic
domains $\mathcal B_\Gamma=I$ and $\bm b_\Gamma=0$, while solid-wall closures
are specified in \cref{sec:walls}.

\subsection{Residual and physical-step acceptance}\label{sec:acceptance}
Evaluating one complete sweep from $\fvec$ without operator splitting defines
the population residual
\begin{equation}\label{eq:residual}
 \fvec_T=T(\fvec),\qquad \bm r=\fvec_T-\fvec,\qquad
 R=\|\bm r\|_{\rm RMS}=
 \left[\frac{1}{QN_\Omega}\sum_{\bm x,k}r_k(\bm x)^2\right]^{1/2}.
\end{equation}
Across all $N_\Omega$ lattice sites, Picard iteration sets $\fvec^+=\fvec_T$,
whereas our corrections alter the update while sharing the same acceptance
criterion. Let $\varepsilon$ denote the residual tolerance and $n_{\min}$ the
prescribed minimum sweep count. Once
$R<\varepsilon$ at the input $\fvec$ of sweep $\ell\ge n_{\min}$, we accept its
sweep image $\fvec_T=T(\fvec)$, reconstruct
$\bm m^{n+1}=\mathcal R_{q,\bm h}(\fvec_T)$ with unchanged history, and then
advance the BDF history.

\Needspace{5\baselineskip}
\section{Conserved-moment structure and the large-step bottleneck}\label{sec:dc}
Unit TRT relaxation removes local zero-moment perturbations in one collision.
We first identify this local structure, then examine how the surviving
conserved modes propagate through the complete sweep.

\subsection{Moment factorization at unit relaxation}\label{sec:moment-structure}
Equation~\eqref{eq:collision-operating} depends on $\fvec$ solely through
$\bm m$, so the local collision derivative factors through the conserved
moments. Throughout the remainder of the paper, $P$ and $E$ also denote their
sitewise block-diagonal extensions on the full lattice.

\begin{proposition}[Moment factorization at unit relaxation]
\label{prop:moment-factorisation}
Fix $a>0$, the histories, and the boundary data, and assume that the boundary
closure is affine. At unit relaxation, let $J^{\mathcal C}$ be the local
collision Jacobian and
$G=\partial_{\bm m}\mathcal C(\bm m;\bm h)$ the collision derivative with
respect to reconstructed moments at fixed history. Then
\begin{equation}\label{eq:local-rank-factorisation}
 J^{\mathcal C}=d_a^{-1}GP,\qquad PG=(1-a/2)I,\qquad
 PJ^{\mathcal C}=\gamma P,\qquad \gamma=\frac{1-a/2}{1+a/2}.
\end{equation}
Thus $\operatorname{rank}J^{\mathcal C}\le d+1$. After streaming and boundary
closure, the complete-sweep Jacobian $L$ at the same state factors as
\begin{equation}\label{eq:global-rank-factorisation}
 L=\mathcal LP,\qquad
 \mathcal L=\mathcal B_\Gamma\mathcal S
 \operatorname{blkdiag}_{\bm x}(G(\bm x)/d_a).
\end{equation}
\end{proposition}
\begin{proof}
Equation~\eqref{eq:collision-operating} gives
$P\mathcal C=(1-a/2)\bm m+\bm h/2$.
Differentiating it and \eqref{eq:macro-reconstruct} at fixed history yields
\eqref{eq:local-rank-factorisation}; composition with streaming and the
affine boundary map yields \eqref{eq:global-rank-factorisation}.
\end{proof}

Every local zero-moment perturbation $\delta\fvec\in\ker P$ is annihilated
by the collision derivative. The remaining error is carried by the
conserved fields and coupled between sites by streaming and boundary closure.

\subsection{Periodic conserved modes and residual scaling}\label{sec:slowmode}
To examine this spatial coupling, consider periodic Fourier modes
$e^{i\kvec\cdot\bm x}$ with streaming symbol
$Z(\kvec)=\operatorname{diag}(e^{-i\kvec\cdot\bm c_k})$.
The following rest-state calculation links slow contraction to the size of
the projected residual.

\begin{proposition}[Periodic conserved-mode reduction and residual scaling]
\label{prop:moment-bottleneck}
Under the fixed-history and unit-relaxation assumptions of
\cref{prop:moment-factorisation}, the following statements hold.
On a periodic domain, linearize at a uniform rest fixed point with
$\bar\uvec=0$ and $\bar{\bm h}=a\bar{\bm m}$. For a population perturbation
$\delta\fvec$, define the raw-moment and reconstructed macroscopic errors by
$\bm e_m=P\delta\fvec$ and $\delta\bm m_R=\bm e_m/d_a$. Their population and
raw-moment Fourier propagation symbols satisfy
\begin{align}
 \widehat{T'}=\gamma ZEP,\qquad
 \mathsf A_m=\gamma PZE,&\qquad
 \det(\lambda I_Q-\widehat{T'})=
 \lambda^{Q-d-1}\det(\lambda I_{d+1}-\mathsf A_m),\label{eq:moment-symbol}\\
 \mathsf A_m(0)=\gamma I,&\qquad
 P\widehat{\bm r}(0)=-(1-\gamma)\widehat{\bm e}_m(0)
 =-a\widehat{\delta\bm m}_R(0),\label{eq:zero-mode-scale}
\end{align}
\end{proposition}
\begin{proof}
At the stated rest state, $\delta\fvec^{\rm eq}=E\delta\bm m$ and
$\delta\bm\Phi=E\delta\bm s=-aE\delta\bm m$, so
$J^{\mathcal C}=\gamma EP$. Streaming supplies $Z$, and Sylvester's
determinant identity gives \eqref{eq:moment-symbol}. Finally,
$Z(0)=I$ and $PE=I$ give \eqref{eq:zero-mode-scale}.
\end{proof}

Conserved errors propagate through the reduced raw-moment symbol
$\mathsf A_m(\kvec)=\gamma PZ(\kvec)E$. As $\dt$ increases ($a\to0^+$), the
Picard contraction factor of the uniform error mode approaches one. For a
fixed reconstructed macroscopic error, the projected residual of this mode
decreases in proportion to $a$.

For that linearized uniform mode with $0<|\gamma|<1$, reducing the error to a
fraction $\theta\in(0,1)$ of its initial value under Picard iteration requires
\begin{equation}\label{eq:zero-mode-work}
 n_0=\left\lceil\frac{\log\theta}{\log|\gamma|}\right\rceil
 \sim\frac{|\log\theta|}{a}
 =\frac{2\dt}{3}|\log\theta|\qquad(a\to0^+,\ \mathrm{BDF2}).
\end{equation}
Since $\log\gamma=-a+O(a^3)$, the required inner-sweep count grows linearly
with $\dt$ and can offset the savings from fewer physical steps over a fixed
integration interval.

This reduction in residual magnitude also affects termination. Because
$\widehat{\delta\bm m}_R(0)=-P\widehat{\bm r}(0)/a$ in
\eqref{eq:zero-mode-scale}, a fixed raw-residual threshold permits $O(1/a)$
uniform moment error as $\dt$ grows. The corrections introduced next target
conserved-moment errors that decay slowly under the baseline iteration.
Physical-step acceptance continues to use the RMS norm of the population
residual, with a tolerance chosen for the physical time step.

\Needspace{5\baselineskip}
\section{Residual correction methods}\label{sec:correctors}
The conserved-moment structure provides a reduced representation of the
linearized defect equation. We use this relation to construct a Fourier
correction on periodic domains and a sparse kinetic correction that includes
the executed solid-wall updates.

\subsection{The defect equation and moment Schur relation}\label{sec:common-correction}
Linearizing $\bm r(\fvec+\delta\fvec)$ around the current iterate and requiring
the first-order residual to vanish yields the ideal Newton defect equation
\begin{equation}\label{eq:correction-equation}
 (I-T')\delta\fvec=\bm r(\fvec).
\end{equation}
Here $T'$ is the Jacobian of $T$ at the current iterate with histories and
boundary data fixed. Replacing it by the factored reference Jacobian
$L=\mathcal LP$
from \eqref{eq:global-rank-factorisation} with $I-L$ nonsingular gives an exact
population--moment Schur equivalence,
\begin{equation}\label{eq:woodbury-schur}
 (I-P\mathcal L)\delta\bm\mu=P\bm r,\qquad
 \delta\fvec=\bm r+\mathcal L\delta\bm\mu,\qquad
 \delta\bm\mu=P\delta\fvec.
\end{equation}
Applying $P$ to $(I-L)\delta\fvec=\bm r$ gives the reduced equation for
$\delta\bm\mu=P\delta\fvec$; substituting it into
$\delta\fvec=\bm r+\mathcal L\delta\bm\mu$ recovers the population increment.
Sylvester's identity ensures simultaneous nonsingularity of $I-\mathcal LP$
and $I-P\mathcal L$.
Combining the population increment with $\fvec+\bm r=T(\fvec)$ gives the
equivalent reference-Jacobian update
\begin{equation}\label{eq:exact-reference-update}
 \fvec^+=T(\fvec)+\mathcal L(I-P\mathcal L)^{-1}P\bm r.
\end{equation}
This expression identifies both the reduced inverse and the lifting action
that an auxiliary correction must approximate.

Equation~\eqref{eq:woodbury-schur} relates population and raw-moment
increments exactly for the same reference operator $L$. The boundary
conditions determine how we approximate the correction in practice. On
periodic meshes, translation invariance permits a $(d+1)$-field Fourier
shadow solve driven by unscaled $P\bm r$, whose lifted increment is added to
$\fvec_T=T(\fvec)$. At solid walls, the sparse population-space operator
$I-L$ retains the ordered bounce-back derivatives, and the increment
$\delta\fvec_{\rm kin}=(I-L)^{-1}\bm r$ is added to $\fvec$.
\Cref{sec:shadow,sec:walls} construct these two updates, \cref{sec:termination}
establishes their fixed-point properties, and \cref{alg:unified} summarizes
their execution under the common acceptance criterion of \cref{sec:acceptance}.

\subsection{Periodic Fourier shadow correction}\label{sec:shadow}
The Fourier symbol of $I-P\mathcal L$ at a uniform rest state on a periodic
lattice is
$I_{d+1}-\mathsf A_m(\kvec)=I_{d+1}-\gamma PZ(\kvec)E$. We approximate the
low-wavenumber correction in \eqref{eq:exact-reference-update} using the
rest-state lift $E$ and a macroscopic acoustic--viscous system. Linearizing
\eqref{eq:macro-pde-pair} at rest, replacing $\partial_t$ with $a$, and using
the Fourier differentiation symbols $i\kvec$ and $-|\kvec|^2$ motivates the
auxiliary symbol
\begin{equation}\label{eq:shadow-operator}
 H(\kvec)=\begin{bmatrix}a&i\kvec^{\mathsf T}\\
 c_s^2i\kvec&(a+\nu|\kvec|^2)I_d\end{bmatrix},\qquad
 \chi(\kvec)=\mathbf1_{\{|k_\alpha|\le\pi/2\ \mathrm{for\ all}\ \alpha\}}.
\end{equation}
For discrete Fourier modes
$\widehat g(\kvec)=\sum_{\bm x}g(\bm x)e^{-i\kvec\cdot\bm x}$ at wrapped
wavenumbers $k_\alpha=2\pi j_\alpha/N_\alpha\in[-\pi,\pi]$, the fixed
mask $\chi(\kvec)$ retains $|k_\alpha|\le\pi/2$ and excludes Nyquist lines.
Writing $\widehat{P\bm r}=(\widehat r_p,\widehat{\bm r}_u)^{\mathsf T}$ and
$b=a+\nu|\kvec|^2$, solving $H(\kvec)\delta\widehat{\bm m}=\widehat{P\bm r}$
for the auxiliary increment in raw-moment space
$\delta\widehat{\bm m}=(\delta\hat p,\delta\uvec)^{\mathsf T}$ gives
\begin{equation}\label{eq:schur}
 \delta\hat p=\frac{\widehat r_p-i\kvec\!\cdot\widehat{\bm r}_u/b}
 {a+c_s^2|\kvec|^2/b},\qquad
 \delta\uvec=\frac{\widehat{\bm r}_u-c_s^2i\kvec\delta\hat p}{b}.
\end{equation}
Applying the mask $\chi$, inverse transforming, and lifting with $E$ adds the
macroscopic shadow correction (MSC) to $T(\fvec)$, defining one inner update of
the periodic macroscopic-shadow-corrected lattice Boltzmann method (MSC-LBM):
\begin{equation}\label{eq:shadow-update}
 \fvec^+=T(\fvec)+E\mathcal F^{-1}
 [\chi H^{-1}\mathcal F(P\bm r)].
\end{equation}
Here $E\mathcal F^{-1}\chi H^{-1}\mathcal F$ approximates
$\mathcal L(I-P\mathcal L)^{-1}$ on $P\bm r$ in
\eqref{eq:exact-reference-update} using $2(d+1)$ scalar FFTs, without dividing
$P\bm r$ by $d_a$ ($d_a^{-1}$ enters at the next reconstruction
$\mathcal R_{q,\bm h}(\fvec^+)$). At $\kvec=\bm0$,
$I-\mathsf A_m(0)=(a/d_a)I$ whereas $H(0)=aI$ (\cref{thm:msc-contraction});
$(I-EP)E=0$ leaves $(I-EP)T(\fvec)$ intact, and conjugate symmetry of $H$ and
$\chi$ preserves real fields.

\Needspace{5\baselineskip}
\begin{theorem}[Zero-mode damping and the axial Nyquist mode]
\label{thm:msc-contraction}
Under the periodic rest-state assumptions of \cref{prop:moment-bottleneck},
with $\nu=c_s^2/2=1/6$, let $\widehat G_{\rm MSC}$ be the corrected population
symbol, $M$ the raw-moment error-propagation matrix, and
$G_{\rm MSC,global}$ the corrected Jacobian on the complete periodic grid. Then
\begin{align}
 \widehat G_{\rm MSC}
 &=\{\gamma ZE+\chi EH^{-1}(\mathsf A_m-I)\}P,\label{eq:msc-pop-symbol}\\
 M&=\mathsf A_m+\chi H^{-1}(\mathsf A_m-I),\qquad
 M(0)=-\frac{a}{2+a}I.\label{eq:msc-moment-symbol}
\end{align}
The nonzero eigenvalues of $\widehat G_{\rm MSC}$ and $M$ coincide.
The zero-mode eigenvalue has magnitude $a/(2+a)$, which is strictly smaller
than $|\gamma|$ exactly when $0<a<1$ (BDF2: $\dt>3/2$), and is $O(a)$ as
$a\to0^+$.
On an even grid, an axial Nyquist mode retains eigenvalue $-\gamma$, so
$\rho(G_{\rm MSC,global})\ge|\gamma|\to1$ as $a\to0^+$.
\end{theorem}
\begin{proof}
For $D=\operatorname{diag}(c_s,I_d)$, the Hermitian part of $DHD^{-1}$ is
$\operatorname{diag}(a,bI_d)>0$. Thus $H$ and $H+I$ are nonsingular.
Linearizing \eqref{eq:shadow-update} and using
$P\widehat{T'}=\mathsf A_mP$ gives \eqref{eq:msc-pop-symbol}.
Its factorization as $UP$, with $PU=M$, proves spectral reduction by
Sylvester's identity. At zero wave number, $H=aI$ and
$\mathsf A_m=\gamma I$, giving
$M(0)=[\gamma+(\gamma-1)/a]I=-aI/(2+a)$.
The comparison $a<|2-a|$ is equivalent to $0<a<1$.
At $\kvec=(\pi,0,\ldots)$, let $\bm e_x$ select the $x$-velocity component in
moment space. Then $\chi=0$ and $ZE\bm e_x=-E\bm e_x$, since the nonzero
components of this velocity lift have $c_{kx}=\pm1$. Hence
$\widehat G_{\rm MSC}E\bm e_x=-\gamma E\bm e_x$.
\end{proof}

Acceleration is thus spectrally selective. In the large-step range $0<a<1$, MSC
maps $\gamma=1-a+O(a^2)$ to $M(0)=-a/(2+a)$, whereas at $a=2$ Picard already
annihilates the zero mode ($\gamma=0$) while $M(0)=-1/2$; because the
uncorrected axial Nyquist mode retains $-\gamma$, nonlinear progress is
monitored directly via the raw residual $R$ (\cref{sec:termination}).

\subsection{Boundary-consistent kinetic correction and factor reuse}\label{sec:walls}
Solid walls couple incoming and reflected populations across boundary links.
Although \eqref{eq:woodbury-schur} permits a moment formulation, we construct
the reference Jacobian $L$ in population space, including streaming and the
ordered bounce-back assignments, and solve the defect equation with reusable
sparse LU factors of $I-L$. Holding $L$ fixed between updates gives a chord
iteration.

With $\mathsf D_k=\bm c_k\bm c_k^{\mathsf T}/c_s^4-I_d/c_s^2$, differentiating
\eqref{eq:collision-operating} at reference moments $\bar{\bm m}$ and fixed
reference history ($\partial\bm s/\partial\bm m=-aI$) gives
\begin{equation}\label{eq:wall-G}
 \begin{aligned}
 G_{k0}&=w_k(1-a/2),\\
 G_{k,u}&=w_k\left[(1-a/2)
 \left(\frac{\bm c_k}{c_s^2}+\mathsf D_k\bar\uvec\right)
 +\tfrac12\mathsf D_k\bar{\bm s}_u\right]^{\mathsf T},
 \qquad J^{\mathcal C}=d_a^{-1}GP.
 \end{aligned}
\end{equation}
The blocks $G_{k0}$ and $G_{k,u}$ evaluate the pressure and velocity
derivatives at reference velocity $\bar\uvec$ and momentum source
$\bar{\bm s}_u$; reference pressure drops out of $G$, while pressure coupling
persists through $G_{k0}$ and the BDF source.

On D2Q9, indexing directions $0$--$8$ as rest, east, north, west, south,
northeast, northwest, southwest, and southeast, each sweep performs modulo
streaming $\mathcal S$ followed by on-grid bounce-back
assignments~\cite{HeZouLuoDembo1997,Ladd1994} in the ordered sequence
\begin{align}
 y=0:&\quad f_2\leftarrow f_4,\quad f_5\leftarrow f_7,\quad f_6\leftarrow f_8,
 \nonumber\\
 y=N_y-1:&\quad f_4\leftarrow f_2,\quad
 f_7\leftarrow f_5-u_{\rm lid}/6,\quad
 f_8\leftarrow f_6+u_{\rm lid}/6,\nonumber\\
 x=0:&\quad f_1\leftarrow f_3,\quad f_5\leftarrow f_7,\quad f_8\leftarrow f_6,
 \nonumber\\
 x=N_x-1:&\quad f_3\leftarrow f_1,\quad f_6\leftarrow f_8,\quad f_7\leftarrow f_5.
 \label{eq:bounceback-rows}
\end{align}
With node counts $N_x,N_y$ and lid speed $u_{\rm lid}$, this
bottom--lid--left--right sequence gives the vertical side walls precedence at
the top corners ($(0,N_y-1)$ for $f_8$, $(N_x-1,N_y-1)$ for $f_7$). Its
derivative applies the same in-place row-copy sequence to the post-streaming
Jacobian; the population-independent terms $\pm u_{\rm lid}/6$ contribute zero.
This yields
\begin{equation}\label{eq:wall-global-jacobian}
 L=\mathcal B_\Gamma\mathcal S
 \operatorname{blkdiag}_{\bm x}(J^{\mathcal C}(\bm x)),\qquad
 (I-L)\delta\fvec_{\rm kin}=\bm r,\qquad
 \fvec^+=\fvec+\delta\fvec_{\rm kin}.
\end{equation}
Unlike \eqref{eq:shadow-update}, $\delta\fvec_{\rm kin}$ updates $\fvec$
directly rather than $T(\fvec)$. Each row of $L$ has at most nine nonzeros
(ten in $I-L$), enabling sequential sparse LU
factorization~\cite{SchenkGartner2004} and triangular factor reuse.

To amortize the sparse factorization cost, both wall-bounded variants factor
$I-L$ once before step~$1$
at the BDF2 rest state $\bar{\bm m}=(1,\bm0)^{\mathsf T}$ with
$\bar{\bm h}_2=a_2\bar{\bm m}$. By pressure independence of $G$, this reference
matches the cavity's $(\hat p,\uvec)=(0,\bm0)$ BDF2 derivative. The frozen
variant reuses the LU factors of this initial reference Jacobian throughout,
including the BDF1 startup step. The nonlinear sweep and residual nevertheless
use the BDF order and history of the current physical step.

The adaptive solver carries the latest factor into the next physical step. At
sweep~$1$ it sets $R_{\rm ref}=R_1$ and resets the consecutive-increase
counter. Two successive increases $R_\ell>R_{\ell-1}$ or
$R_\ell>10R_{\rm ref}$ trigger reconstruction from the current iterate,
matrix assembly and factorization at the current step's BDF order and unchanged
history, and counter resets with $R_{\rm ref}=R_\ell$. At most three refreshes
are allowed per step; iteration then continues with the latest factor.
Provided $I-L$ remains nonsingular, reuse preserves the fixed-point set.

\subsection{Fixed-point preservation and termination}\label{sec:termination}
\Needspace{5\baselineskip}
\begin{proposition}[Preservation of the discrete fixed-point set]
\label{prop:fixed-point-preservation}
Fix the current physical-step histories and boundary data. If
$a,\nu,c_s^2>0$ and each wall matrix $I-L$ used is nonsingular, both
\eqref{eq:shadow-update} and \eqref{eq:wall-global-jacobian} have exactly
the fixed points of $T$. The reference matrix may come from an earlier state.
\end{proposition}
\begin{proof}
If $\bm r=0$, both corrections vanish and $T(\fvec)=\fvec$.
Conversely, suppose that $\fvec$ is a fixed point of the periodic corrected
update. Applying $P$ to its fixed-point equation and taking the Fourier
transform gives
$-\widehat{P\bm r}=\chi H^{-1}\widehat{P\bm r}$.
Outside the mask this gives $\widehat{P\bm r}=0$; inside it gives
$(H+I)\widehat{P\bm r}=0$. The positivity argument in
\cref{thm:msc-contraction} proves $H+I$ nonsingular. The correction
therefore vanishes everywhere and the full fixed-point equation gives
$\bm r=0$. For the wall update, a fixed point has
$(I-L)^{-1}\bm r=0$, hence $\bm r=0$.
\end{proof}

\Cref{prop:fixed-point-preservation} establishes equality of the
discrete fixed-point sets. The periodic analysis characterizes the linearized
rest-state behavior, and the numerical experiments assess convergence in the
nonlinear runs.
\Cref{alg:unified} evaluates the population residual \eqref{eq:residual} before
applying either correction.
When the tolerance and minimum sweep count are satisfied, it accepts the sweep
image, reconstructs the macroscopic state with unchanged history, and then
advances the history. Otherwise, the correction supplies the next sweep's
input; failure to satisfy the acceptance criterion within the iteration limit
causes step rejection.
\begin{algorithm}[!htbp]
\caption{Residual correction at one physical time step.}\label{alg:unified}
\small
\begin{algorithmic}[1]
\Require current $q,a_q,\bm h_q,\Gamma$; initial $\fvec$; $\varepsilon,n_{\min},K_{\max}$
\State keep histories and boundary data fixed; initialize wall refresh counters
\For{$\ell=1,\ldots,K_{\max}$}
 \State $\fvec_T\gets T_{q,\bm h_q,\Gamma}(\fvec)$;
 $\bm r\gets\fvec_T-\fvec$; $R_\ell\gets\|\bm r\|_{\rm RMS}$
 \If{$R_\ell<\varepsilon$ and $\ell\ge n_{\min}$}
  \State accept $\fvec^{n+1}\gets\fvec_T$ and
  $\bm m^{n+1}\gets\mathcal R_{q,\bm h_q}(\fvec_T)$; advance history; \Return
 \EndIf
 \If{periodic shadow correction}
  \State $\fvec\gets\fvec_T+E\mathcal F^{-1}[\chi H^{-1}\mathcal F(P\bm r)]$
 \Else
  \State refresh and factor $A=I-L$ if triggered by \cref{sec:walls}
  \State solve $A\delta\fvec=\bm r$; $\fvec\gets\fvec+\delta\fvec$
 \EndIf
\EndFor
\State \Return nonconvergence; do not accept the physical step
\end{algorithmic}
\end{algorithm}

\FloatBarrier
\section{Numerical experiments}
\label{sec:results}
The numerical experiments evaluate discrete consistency, temporal accuracy,
inner-iteration convergence, and runtime performance across two- and
three-dimensional periodic Taylor--Green vortex (TGV) decay and the impulsively
started two-dimensional lid-driven cavity.

\subsection{Test problems, error measures, and computational settings}
\label{sec:protocol}
Nominal tests use lattice units $\Delta x=\Delta t_{\rm ac}=1$, $\nu=1/6$, and
$\Lambda=1/4$ ($\omega^+=\omega^-=1$), with $\Re=UN/\nu$ and $\Ma=U/\cs$ for
directional node count $N$ and characteristic speed $U$. In 2D periodic TGV
decay (wavenumber $2\pi/N$, amplitude $U=\Re\nu/N$, equilibrium initialization
with analytic pressure \eqref{supp:eq:tgv2d-pressure-init}, horizon
$T_{\rm end}=\operatorname{nint}(0.15N^2)$), the absolute endpoint error is the
velocity RMS difference from the analytic field, and dividing by the numerical
endpoint velocity RMS (floored at $10^{-12}$) gives the relative error. In 3D
D3Q19 TGV decay ($T_{\rm end}=\operatorname{nint}(0.075N^2)$, zero initial
pressure perturbation and solenoidal velocity \eqref{supp:eq:tgv3d-ic}), the
$1\%$ criterion normalizes the endpoint velocity RMS difference from the
same-grid explicit LBM solution by $U/2$, whereas matched accuracy compares RMS
departures of the explicit solution and the solution obtained with dual-time
stepping from the diagnostic modal decay field
$\uvec(\bm x,0)\exp(-4\nu k_0^2t)$. The enclosed cavity starts from rest with
lid speed $U=u_{\rm lid}=\Re\nu/N$ over
$T_{\rm end}=\operatorname{nint}(1.5N^2)$ and normalizes the RMS difference
between the endpoint velocity fields obtained with dual-time stepping and with
the explicit baseline by $u_{\rm lid}$ \eqref{supp:eq:cavity-error}.

Full candidate sets of $(N,\Re,S,\varepsilon)$ with $\dt=T_{\rm end}/S$ appear
in \cref{supp:tab:protocol,supp:sec:protocol,supp:sec:full-results}: 2D TGV
spans $N=32$--$1024$ ($\Re\in\{1,100\}$) and $\Re=10^{-4}$--$10^2$ ($N=256$)
with $K_{\max}=300$; 3D TGV spans $N\in\{32,64,128\}$ ($\Re\in\{1,100\}$,
$K_{\max}=300$); and cavity timing tests use $N\in\{64,128\}$
($\Re\in\{10,100\}$, $K_{\max}=4000$, plus $N\in\{32,256\}$ for single-step
tests). A physical step is accepted once $R_\ell<\varepsilon$ after at least
$n_{\min}=2$ sweeps in 2D periodic runs and $n_{\min}=1$ in 3D and cavity runs;
nonconvergence within $K_{\max}$ on any step rejects the trajectory. At
$t=T_{\rm end}$, matched accuracy requires absolute error within three times the
explicit baseline, while relative criteria require errors within $1\%$ (and
$0.1\%$ in the cavity). Periodic benchmarks evaluate MSC-LBM
(\cref{sec:shadow})---against Picard-smoothed FAS W-cycle multigrid (MG) and
weighted-Jacobi FAS multigrid (Gsell-type; \cref{supp:app:comparison-mg}) in 2D
and against Picard in 3D---whereas cavity benchmarks compare Picard with the
frozen and adaptive reference kinetic corrections (\cref{sec:walls}).
Each entry in \cref{tab:tgv2d,tab:tgv3d,tab:cavity} selects $(S,\varepsilon)$
minimizing a prescribed work proxy $W$ (weighted fine/coarse sweeps and spectral
solves in periodic runs, kinetic sweeps in the cavity;
\cref{supp:sec:parameter-range}) rather than elapsed time; fixed-step tests fix
$(S,\varepsilon)$ directly.

All timings run serially on one pinned physical core of an exclusive dual-socket
Intel Xeon Platinum 8358 node, taking the minimum over two explicit runs and
four runs with dual-time stepping (startup and stopping details in
\cref{supp:sec:comparison-protocols}). For periodic flows,
$\Swall=t_{\rm explicit}/t_{\rm scheme}$ times the full time-stepping loop,
excluding allocation and terminal error evaluation. For the cavity,
$\Smarch=t_{\rm explicit,march}/t_{\rm scheme,march}$ times physical marching
(sweeps, triangular back-solves, and adaptive refactorizations) after pre-march
setup (a preliminary two-step Picard pass used only when preparing a nonzero
reference state, rest-state assembly and LU factorization of $I-L$, and reset to
rest). Fixed-step comparisons in \cref{supp:sec:total-runtime} also report
$\Stotal=t_{\rm explicit,total}/t_{\rm scheme,total}$ across four repetitions of
both the explicit baseline and the solver with dual-time stepping, enclosing all
initialization, factorization, and marching.

\subsection{Consistency and temporal accuracy}
\Cref{prop:fixed-point-preservation} guarantees identical discrete fixed-point
sets under its stated conditions. On one interior BDF2 step of 2D TGV decay from
analytic histories at $t=0$ and $\dt$
($N=128,\Re=100,\dt=38.9,\varepsilon=10^{-12},K_{\max}=600$), MSC-LBM and Picard
velocities differ by $3.57\times10^{-11}$ in vector RMS ($4.15\times10^{-10}$
relative). Over two-step (BDF1--BDF2) runs at $\Re=1$, tightening $\varepsilon$
from $10^{-12}$ to $10^{-14}$ reduces full-population RMS differences
proportionally from $5.27\times10^{-11}$ to $5.27\times10^{-13}$ in 2D
($N=128,\dt=100$) and from $4.61\times10^{-12}$ to $4.25\times10^{-14}$ in 3D
($N=32,\dt=20$; \cref{supp:sec:verification}). Directional
finite-difference checks across BDF1/BDF2 cavity states confirm that $I-L$
matches the executed bounce-back sweep to roundoff once the exact quadratic
remainder is removed (\cref{supp:sec:verification}), and independent cavity
runs (\cref{supp:tab:verification}) agree to all three displayed digits at
$(64,10)$ and $(64,100)$ and within $2\%$ at $(128,100)$
($4.82\times10^{-7}$--$4.91\times10^{-7}$).

Halving $\dt$ on fixed periodic meshes with MSC-LBM at $\varepsilon=10^{-12}$
and $K_{\max}=20000$ (terminal residuals $\le9.57\times10^{-13}$;
\cref{supp:sec:verification}) yields observed orders of $2.178$ and $2.062$ in
2D and $2.111$ and $2.069$ in 3D (\cref{tab:temporal}), confirming second-order
temporal self-convergence of the slowly evolving velocity fields.

\begin{table}[!tbp]
\centering
\caption{Temporal self-convergence of periodic velocity at fixed resolution for
2D $(N,\Re,T_{\rm end})=(128,1,1600)$ and 3D $(32,1,160)$, with component-wise
terminal velocity RMS difference $D_i$ \cref{supp:eq:observed-order} and
$p_i=\log_2(D_i/D_{i+1})$.}
\label{tab:temporal}
\small
\begin{tabular}{lrr@{\qquad}lrr}
\toprule
\multicolumn{3}{c}{2D TGV} & \multicolumn{3}{c}{3D TGV}\\
$\dt_i,\dt_i/2$ & $D_i$ & $p_i$ & $\dt_i,\dt_i/2$ & $D_i$ & $p_i$\\
\midrule
$200,100$ & $1.3655\times10^{-6}$ & 2.178 & $16,8$ & $1.7198\times10^{-6}$ & 2.111\\
$100,50$  & $3.0181\times10^{-7}$ & 2.062 & $8,4$  & $3.9826\times10^{-7}$ & 2.069\\
$50,25$   & $7.2263\times10^{-8}$ & --    & $4,2$  & $9.4932\times10^{-8}$ & --\\
\bottomrule
\end{tabular}
\end{table}

\subsection{Two-dimensional periodic flow}
\label{sec:periodic_results}
\begin{figure}[!tbp]
\centering
\includegraphics[width=\textwidth]{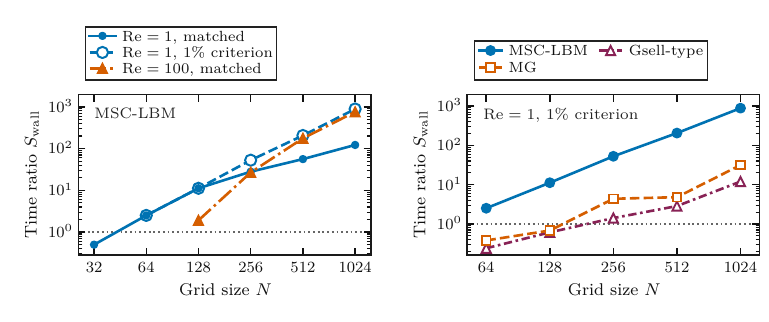}
\caption{Two-dimensional periodic speedups (\cref{tab:tgv2d}). Left: MSC-LBM
under the matched-accuracy and $1\%$ criteria. Right: MSC-LBM versus FAS
multigrid under the $1\%$ criterion at $\Re=1$, where the dotted line marks
equal runtime ($\Swall=1$). No $N=32$ run meets the $1\%$ criterion; concentric
markers at $N=64,128$ ($\Re=1$) reflect identical selected runs, and $\Re=100$
starts at $N=128$ (\cref{supp:tab:protocol}).}
\label{fig:periodic2d}
\end{figure}

At $N=128$ and $\dt=38.9$, Picard tail contraction
$\rho_{\mathrm{obs}}=0.958$--$0.960$ \cref{supp:eq:rho-observed} at $\Re=1$
and $100$ matches the analytical rest-state limit $\gamma_\dt=0.9622$, whereas
$W$-selected MSC-LBM configurations average $2.0$--$3.2$ sweeps per step across
all tested grids. Each periodic correction pairs a $Q$-population sweep with
$2(d+1)$ complex FFTs, Schur solves, and lifting; runtime ratios grow once
outer-step savings amortize this cost (\cref{supp:sec:parameter-range}). At
$(N,\Re)=(256,1)$, MSC-LBM reaches $\Swall=27.95$ under matched accuracy and
$52.71$ under the $1\%$ criterion, compared with $0.47$--$0.87$ and
$1.41$--$4.37$ for the two FAS multigrid solvers
(\cref{tab:tgv2d,fig:periodic2d}).

\begin{table}[!tbp]
\centering
\caption{Two-dimensional Taylor--Green vortex: explicit-to-scheme time ratio
$\Swall$ under the matched-accuracy and $1\%$ relative-error criteria
(\cref{sec:protocol}). Explicit runtimes are in seconds; a dash indicates that
no tested configuration satisfied the accuracy criterion.}
\label{tab:tgv2d}
\small
\begin{tabular}{lrrrr}
\toprule
\multicolumn{5}{c}{\textbf{matched-accuracy criterion}, $\Re=1$}\\
$N$ & explicit (s) & MG & Gsell-type & MSC-LBM\\
\midrule
32   & 0.006 & 0.32 & 0.19 & 0.50\\
64   & 0.098 & 0.38 & 0.29 & 2.51\\
128  & 1.613 & 0.68 & 0.31 & 11.20\\
256  & 26.78 & 0.47 & 0.87 & \textbf{27.95}\\
512  & 434   & --   & --   & 55.98\\
1024 & 7576  & --   & --   & 122.02\\
\midrule
\multicolumn{5}{c}{\textbf{matched-accuracy criterion}, $\Re=100$}\\
$N$ & explicit (s) & MG & Gsell-type & MSC-LBM\\
\midrule
128  & 1.615 & 0.66 & 0.73 & 1.91\\
256  & 26.78 & 1.54 & 1.14 & 26.81\\
512  & 433.9 & 1.16 & 2.49 & 175.47\\
1024 & 7575  & 2.74 & 7.62 & 755.86\\
\midrule
\multicolumn{5}{c}{\textbf{$1\%$ relative-error criterion}, $\Re=1$}\\
$N$ & explicit (s) & MG & Gsell-type & MSC-LBM\\
\midrule
32   & 0.006 & --    & --    & --\\
64   & 0.098 & 0.38  & 0.24  & 2.51\\
128  & 1.613 & 0.68  & 0.61  & 11.20\\
256  & 26.78 & 4.37  & 1.41  & \textbf{52.71}\\
512  & 434   & 4.80  & 2.85  & 205.68\\
1024 & 7576  & 31.47 & 11.98 & 885.98\\
\bottomrule
\end{tabular}
\end{table}

Refining $N$ from $64$ to $1024$ at fixed $\Re$ raises explicit runtime from
$0.098\,\mathrm{s}$ to $7576\,\mathrm{s}$ under $O(N^4)$ diffusive scaling while
MSC-LBM stays near two to three sweeps per step: $\Swall$ rises from $2.51$ to
$122.02$ under matched accuracy and to $885.98$ under the $1\%$ criterion at
$\Re=1$, and from $1.91$ to $755.86$ across $N=128$--$1024$ at $\Re=100$. At
$N=32$ ($0.006\,\mathrm{s}$ explicit over $T_{\mathrm{end}}=154$), FFT overhead
gives $\Swall=0.50$ under matched accuracy and no tested step meets the $1\%$
criterion. Including setup when starting directly from
\eqref{supp:eq:tgv2d-pressure-init} at $\Re=1$ and $\varepsilon=10^{-6}$,
MSC-LBM attains $\Stotal=7.89,15.93,29.55$ ($S=16,32,64$ under matched accuracy)
and $\Stotal=13.91,60.25,222.81$ ($S=8$ under the $1\%$ criterion) on
$N\in\{128,256,512\}$ (\cref{supp:tab:total-runtime}).

\subsection{Three-dimensional periodic flow}
\label{sec:transfer3d}
\begin{figure}[!tbp]
\centering
\includegraphics[width=\textwidth]{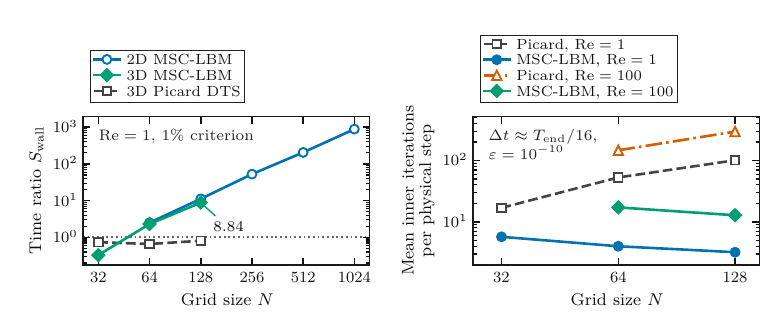}
\caption{Grid dependence of periodic acceleration. Left: 2D and 3D time ratios
under the $1\%$ criterion ($N=32$ is absent in 2D, and 3D $\Re=100$ starts at
$N=64$; \cref{tab:tgv3d}). Right: mean 3D inner sweeps per step at
$\dt\approx T_{\mathrm{end}}/16$ and $\varepsilon=10^{-10}$, where Picard at
$(128,1)$ and $\Re=100$ includes steps capped at $K_{\max}=300$.}
\label{fig:transfer3d}
\end{figure}

On D3Q19, MSC-LBM solves the $4\times4$ Fourier system \eqref{eq:schur} on
$d+1=4$ conserved fields and lifts via \eqref{eq:linear-lift} to $Q=19$
populations under unchanged unit TRT relaxation and mask $|k_\alpha|\le\pi/2$.
At prescribed $\dt\approx T_{\mathrm{end}}/16$, $\varepsilon=10^{-10}$, and
$K_{\max}=300$ (\cref{fig:transfer3d}, right), refining $N$ from $32$ to $128$
($\dt=\lfloor T_{\mathrm{end}}/16\rfloor$ from $4$ to $76$) lowers MSC-LBM mean
sweeps per step from $5.7$ to $3.2$ ($\Re=1$) and from $17.2$ to $12.8$
($\Re=100$), whereas Picard jumps from $16.9$ to $100.8$ ($\Re=1$) and reaches
$295.8$ at $(128,100)$ with initial steps capped at $K_{\max}$. Under exact
$S=16$, Picard hits $K_{\max}$ on step~$1$ at $(128,1)$ while MSC-LBM averages
$3.2$ sweeps per step, and at $(64,1)$ ($\dt=19.19$, all steps convergent)
MSC-LBM cuts mean sweeps from $54.6$ ($87.85\,\mathrm{s}$) to $4.1$
($16.32\,\mathrm{s}$, or $5.38$ times faster at the same $\dt$).

\begin{table}[!tbp]
\centering
\caption{Three-dimensional Taylor--Green vortex (D3Q19): explicit-to-scheme
time ratio $\Swall$ under matched-accuracy (modal reference) and $1\%$
relative-error (same-grid explicit reference) criteria (\cref{sec:protocol}).
Dashes mark $(32,100)$ (untested) or cases with no qualifying configuration.}
\label{tab:tgv3d}
\small
\begin{tabular}{lrrr@{\qquad}rr}
\toprule
 & \multicolumn{3}{c}{$\Re=1$} & \multicolumn{2}{c}{$\Re=100$}\\
$N$ & explicit (s) & Picard & MSC-LBM & Picard & MSC-LBM\\
\midrule
\multicolumn{6}{l}{\emph{matched-accuracy criterion}}\\
32  & 0.452 & 1.00 & 1.07 & -- & --\\
64  & 14.44 & 9.05 & 5.35 & 0.23 & 0.35\\
128 & 464.7 & 21.11 & \textbf{27.85} & 0.41 & 2.84\\
\midrule
\multicolumn{6}{l}{\emph{$1\%$ relative-error criterion}}\\
32  & 0.452 & 0.74 & 0.33 & -- & --\\
64  & 14.44 & 0.66 & 2.31 & -- & --\\
128 & 464.7 & 0.81 & \textbf{8.84} & 0.41 & 2.84\\
\bottomrule
\end{tabular}
\end{table}

Net 3D ratios trade saved sweeps against eight complex $N^3$ FFTs per spectral
correction over the shorter $0.075N^2$ horizon (\cref{tab:tgv3d}). Because the
two 3D criteria use different reference fields, neither relaxes the other: at
$(64,1)$, Picard exceeds MSC-LBM under modal matched accuracy ($\Swall=9.05$
versus $5.35$) where few sweeps are needed, whereas under the same-grid $1\%$
criterion MSC-LBM reaches $2.31$ versus $0.66$. At $N=128$ ($464.7\,\mathrm{s}$
over $1229$ explicit steps), MSC-LBM attains $\Swall=27.85$ under matched
accuracy and $8.84$ under the $1\%$ criterion at $\Re=1$, compared with $21.11$
and $0.81$ for Picard, and reaches $2.84$ versus $0.41$ at $\Re=100$. Including
setup at fixed $(S,\varepsilon)=(16,10^{-10})$ and $\Re=1$ under the $1\%$
criterion, $\Stotal$ rises from $0.89$ at $N=64$ to $4.45$ at $N=128$
($104.3\,\mathrm{s}$ versus $464.5\,\mathrm{s}$ explicit;
\cref{supp:tab:total-runtime}).

\subsection{Enclosed cavity flow}
\label{sec:cavity}
\begin{figure}[!tbp]
\centering
\includegraphics[width=\textwidth]{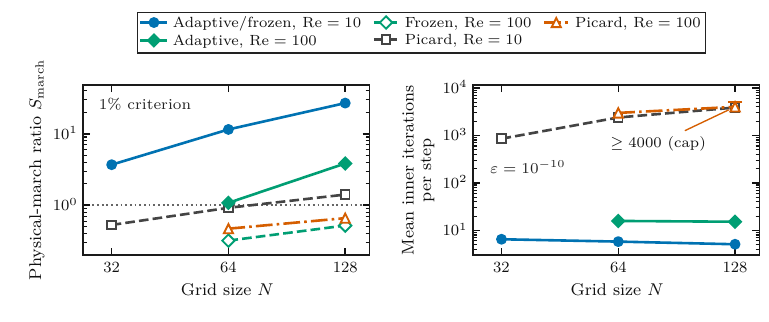}
\caption{Cavity acceleration. Left: physical-marching ratios $\Smarch$ under the
$1\%$ criterion (the shared $\Re=10$ curve uses refresh-free adaptive timings on
$N=64,128$ plus the $N=32$ diagnostic point, and $\Re=100$ starts at $N=64$;
\cref{tab:cavity}). Right: mean inner sweeps per step at $\dt=T_{\mathrm{end}}/8$
and $\varepsilon=10^{-10}$, where frozen $\Re=100$ diverges and Picard for
$N\ge64$ includes steps capped at $K_{\max}=4000$, saturating on all steps at
$(128,100)$.}
\label{fig:cavity}
\end{figure}

In the impulsively started lid-driven cavity~\cite{Ghia1982} over
$T_{\mathrm{end}}=\operatorname{nint}(1.5N^2)$, inner convergence hinges on
discrete bounce-back compatibility and reference-state tracking
(\cref{tab:cavity,fig:cavity}). On step~$3$ at $(N,\Re,\dt)=(64,100,10)$, after
Picard preprocessing from BDF1/BDF2 histories to $R=8.18\times10^{-6}$, one
nodal continuum-Stokes correction with homogeneous Dirichlet wall velocities
amplifies $R$ $63.8$-fold to $5.21\times10^{-4}$, concentrating $99.89\%$ of the
squared residual on wall nodes ($6.15\%$ of the grid); from the same state, one
rest-state kinetic correction via $I-L$ lowers $R$ to $5.22\times10^{-6}$
($0.638$ of the value before correction). On step~$3$ at $(64,10,38.9)$ and
$\varepsilon=10^{-10}$, the bounce-back Jacobian correction converges in $5$
iterations versus $280$ Picard sweeps.

At $\Re=100$, advection shifts the flow away from rest: on step~$3$ at $N=64$
($\dt=38.9$, $\varepsilon=10^{-10}$), evaluating
\eqref{eq:wall-G}--\eqref{eq:wall-global-jacobian} at the accepted step-$2$
macroscopic field (an Oseen-type nonzero-velocity reference) rather than at rest
cuts $\rho_{\mathrm{obs}}$ from $0.342$ to $0.053$ and iterations from $15$ to
$7$, compared with $334$ Picard sweeps, converging in $8,7,5,4$ sweeps with
$\rho_{\mathrm{obs}}=0.078,0.053,0.025,0.011$ across $N\in\{32,64,128,256\}$ as
$\Ma$ decreases and motivating the adaptive updates in \cref{sec:walls}. Over
full trajectories at $\varepsilon=10^{-10}$, the eight-step $(64,10)$ run
($\dt=768$) averages $5.8$ sweeps per step while Picard hits $K_{\max}=4000$ on
step~$1$ ($2389.5$ mean sweeps); at $\dt=384$, where Picard converges on every
step, mean sweeps drop from $1049.2$ to $4.4$ at $(64,10)$ ($S=16$) and from
$503.7$ to $2.4$ at $(128,10)$ ($S=64$).

\begin{table}[!tbp]
\centering
\caption{Two-dimensional enclosed cavity: physical-marching time ratio $\Smarch$
under $1\%$ and $0.1\%$ criteria (\cref{sec:protocol}), excluding pre-march
setup and initial rest-state LU factorization while including triangular
back-solves and adaptive refactorizations. Frozen $\Re=100$ uses smaller steps
as larger steps diverge; dashes denote nonconvergence within $K_{\max}$.}
\label{tab:cavity}
\small
\begin{tabular}{llrrrr}
\toprule
criterion & case $(N,\Re)$ & explicit (s) & Picard & kinetic frozen & kinetic adaptive\\
\midrule
$1\%$   & $(64,10)$   & 0.862 & 0.92 & 11.49 & 11.49\\
$1\%$   & $(128,10)$  & 14.72 & 1.41 & 26.96 & \textbf{26.86}\\
$1\%$   & $(64,100)$  & 0.859 & 0.47 & 0.32  & 1.08\\
$1\%$   & $(128,100)$ & 14.75 & 0.66 & 0.52  & \textbf{3.84}\\
\midrule
$0.1\%$ & $(64,10)$   & 0.862 & 0.35 & 6.16  & 6.11\\
$0.1\%$ & $(128,10)$  & 14.72 & --   & 18.68 & 18.68\\
$0.1\%$ & $(64,100)$  & 0.859 & 0.47 & 0.32  & 0.86\\
$0.1\%$ & $(128,100)$ & 14.75 & --   & 0.52  & 2.99\\
\bottomrule
\end{tabular}
\end{table}

At $\Re=10$, the rest-state factor converges throughout without refresh, so
frozen and adaptive runs yield bitwise-identical fields ($\Smarch=11.49$--$26.96$
under the $1\%$ criterion and $6.11$--$18.68$ under the $0.1\%$ criterion,
whereas Picard hits $K_{\max}=4000$ at $N=128$ under the $0.1\%$ criterion). At
$\Re=100$, reference updates govern both efficiency and large-step stability: on
$N=128$ at $\dt=384$ ($\varepsilon=10^{-10}$), one refresh cuts mean sweeps from
$11.0$ to $6.6$ and total solver time by $25\%$, while at larger steps
($\dt\ge768$ on $N=64$, $\dt\ge1536$ on $N=128$) frozen iteration diverges
($\Smarch=0.32$--$0.52$ at smaller steps) whereas adaptive relinearization
reaches $\Smarch=1.08$ and $3.84$ under the $1\%$ criterion and $2.99$ at
$N=128$ under the $0.1\%$ criterion as $\Ma$ drops from $0.451$ ($N=64$) to
$0.226$ ($N=128$). At $(128,100)$ under the $1\%$ criterion, $\Swork=472$ versus
$\Smarch=3.84$ because $W$ counts kinetic sweeps alone whereas marching also
executes sparse LU back-solves on the $9N^2$-unknown system and adaptive
refactorizations. At $(128,10)$ with $(S,\varepsilon)=(64,10^{-10})$ ($\dt=384$;
\cref{supp:tab:total-runtime}), marching takes $3.67\,\mathrm{s}$ ($\Smarch=4.09$
versus $14.97\,\mathrm{s}$ explicit) and pre-march setup takes $9.22\,\mathrm{s}$
($7.22\,\mathrm{s}$ preliminary Picard pass, $1.97\,\mathrm{s}$ initial LU,
$0.030\,\mathrm{s}$ CSR assembly), yielding $\Stotal=1.16$ over explicit stepping
and a $3.50$-fold total speedup over Picard ($45.07\,\mathrm{s}$).

\subsection{Influence of the physical time step and relaxation parameters}
\label{sec:range-main}
\begin{figure}[!tbp]
\centering
\includegraphics[width=\textwidth]{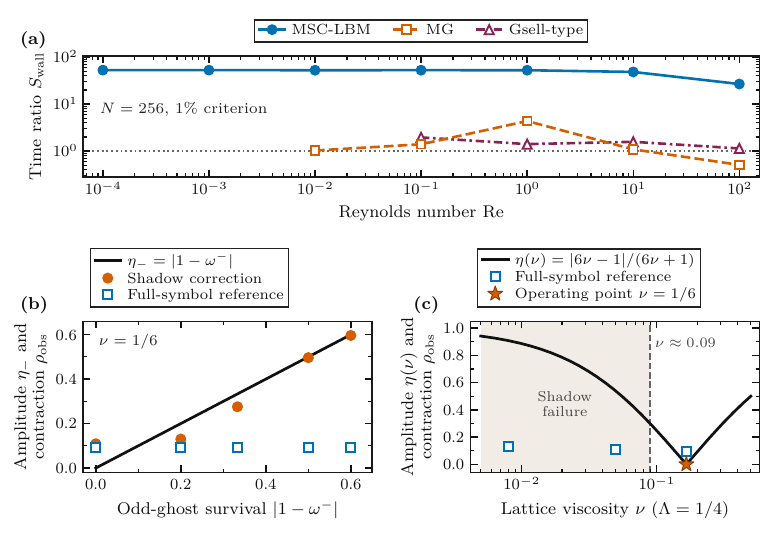}
\caption{Parameter dependence of the correction. (a) Time ratio across six
decades of $\Re$ at $N=256$ under the $1\%$ criterion, where missing points
indicate that no tested configuration met the criterion. (b) Contraction versus
$\Lambda$ at $\nu=1/6$ with the analytic odd-mode amplitude. (c) Analytic
$\eta(\nu)$ law, shadow failure region, and full-symbol reference along
$\Lambda=1/4$, where markers show periodic measurements and curves show analytic
amplitudes.}
\label{fig:validity}
\end{figure}

At unit TRT relaxation on the $N=256$ periodic grid under the $1\%$ criterion,
MSC-LBM maintains $\Swall=26.81$--$52.92$ across $\Re=10^{-4}$--$10^2$
(\cref{fig:validity}a), with fixed-step $(S,\varepsilon)=(8,10^{-6})$ total
ratios $\Stotal=60.23,60.25,37.63$ at $\Re\in\{0.1,1,100\}$ when initialized
from \eqref{supp:eq:tgv2d-pressure-init} (\cref{supp:tab:total-runtime}). At
$(N,\Re,T_{\mathrm{end}})=(128,100,2458)$, MSC-LBM converges at $S=32$
($\dt\approx76.8$) and diverges at $S=16$ ($\dt\approx153.6$); across
$\Re\in\{120,150,180\}$ ($N=128$) and $\Re\in\{250,300,360\}$ ($N=256$), all six
cases converge at $\dt\approx19.2$ and fail at $\dt\approx76.8,153.6$, whereas
refining $N$ at fixed $\Re$ ($\Ma\propto N^{-1}$) admits $\dt\approx2\times10^4$
at $N=1024$. At $(256,1)$ and $(S,\varepsilon)=(8,10^{-6})$ ($\dt=1228.75$), the
$O(a)$ residual attenuation in \eqref{eq:zero-mode-scale} lets Picard and FAS
multigrid satisfy $R_\ell<10^{-6}$ in $1$--$2$ iterations per step while leaving
$85.7\%$ and $1293\%$ relative velocity errors, whereas MSC-LBM inverts the
low-wavenumber moment defect to reach $0.836\%$ in two sweeps per step
(\cref{supp:sec:residual-scaling}). Varying $S\in\{8,32,64\}$ at
$\varepsilon=10^{-6}$ gives non-monotone errors ($0.836\%,0.00811\%,0.0175\%$)
once algebraic stopping error overtakes BDF error at $S=64$; tightening
$\varepsilon$ to $10^{-10}$ lowers the $S=64$ error to $0.00235\%$, while $S=32$
stalls near $1.1\times10^{-10}$ at $K_{\max}=300$ and is rejected
(\cref{supp:tab:residual-tolerance}).

Away from unit relaxation, we compare MSC-LBM on one interior BDF2 step
($N=128,\varepsilon=10^{-10}$) against a nine-field D2Q9 Fourier reference
solver (requiring three times as many FFTs) with rest-state collision Jacobian
$J$ evaluated by forward differences (step $10^{-7}$) at $(\hat p,\uvec)=(1,\bm0)$
and $\widehat L(\kvec)=Z(\kvec)J$:
\begin{equation}
 [I-\widehat L(\kvec)]\,\delta\widehat\fvec
 =\widehat{\bm r}(\kvec),\qquad
 \fvec^{(\ell+1)}=\fvec^{(\ell)}+\delta\fvec.
 \label{eq:full-symbol}
\end{equation}
Here $\delta\fvec=\mathcal F^{-1}\delta\widehat\fvec$ with
$\delta\widehat\fvec\in\mathbb C^9$. With parity projectors
$(S_\pm g)_k=(g_k\pm g_{\bar k})/2$ at a uniform rest fixed point
($\bar\uvec=\bm0,\bar{\bm h}=a\bar{\bm m}$, history fixed), the general TRT
collision Jacobian splits into $\operatorname{diag}(w_k)^{-1}$-orthogonal
conserved and zero-moment sectors,
\begin{equation}
 J=K_\Omega+\gamma EP,\qquad
 K_\Omega=(I_Q-\omega^+S_+-\omega^-S_-)(I_Q-EP),\qquad
 \gamma=\frac{1-a/2}{1+a/2}.
 \label{eq:general-trt-jacobian}
\end{equation}
Since $PK_\Omega=0$ and $K_\Omega E=0$, $K_\Omega$ acts strictly on $\ker P$
while $PS_\pm E$ cancels $\omega^\pm$ in $PJ$, leaving $\gamma$ unchanged and
amplifying even/odd nonconserved (ghost) modes by $1-\omega^\pm$. Along
$\Lambda=1/4$, \eqref{eq:trt} gives $\tau^+=1/2+3\nu$ and
$\tau^-=1/2+1/(12\nu)$, so
\begin{equation}
 |1-\omega^+|=|1-\omega^-|
 =\eta(\nu)=\frac{|6\nu-1|}{6\nu+1}.
 \label{eq:eta}
\end{equation}
Here $\eta(\nu)\in[0,1)$ measures post-collision ghost retention before
streaming couples nonconserved modes back into conserved moments
($PZ(\kvec)K_\Omega\ne0$ for $\kvec\ne\bm0$). Periodic scans confirm this
mechanism (\cref{fig:validity}b,c): varying $\Lambda$ at $\nu=1/6$ activates
$|1-\omega^-|>0$ and raises MSC-LBM contraction from $0.109$ to $0.596$ without
divergence, whereas varying $\nu$ along $\Lambda=1/4$ at $\Ma=0.1$ ($\dt=38.9$)
raises $\rho_{\mathrm{obs}}$ to $0.530$ ($\nu=0.10,\eta=0.250$) and $0.716$
($\nu=0.095,\eta=0.274$) before divergence at $\nu=0.09$ ($\eta=0.299$), while
\eqref{eq:full-symbol} maintains $\rho_{\mathrm{obs}}=0.091$--$0.131$,
motivating low-cost subspace enrichment away from unit relaxation.

Time-scale separation defines the favorable regime: at $\nu=1/6$,
$t_\nu\sim6N^2$ and $t_c\sim6N^2/\Re$ require $O(6N^2/\max(1,\Re))$ explicit
steps over $\min(t_\nu,t_c)$, providing wide headroom for large physical steps
on fine low-Mach meshes while accuracy, nonlinear stability, and auxiliary cost
account for the $N=32$ periodic loss and moderate $\Re=100$ cavity gains.

\section{Conclusions}\label{sec:conclusions}
The conserved-moment structure of the complete discrete sweep provides a
basis for accelerating inner iterations in lattice Boltzmann methods with
dual-time stepping. At unit TRT
relaxation, collision depends on incoming populations solely through $d+1$
reconstructed moments, while periodic rest-state analysis reveals weak damping
of the uniform conserved mode at large $\dt$ together with $O(\dt^{-1})$
attenuation of its projected residual. These properties motivate the periodic
Fourier macroscopic-shadow-corrected lattice Boltzmann method (MSC-LBM) and a
sparse reference kinetic correction retaining executed bounce-back updates at
solid walls, both preserving the discrete fixed-point set under
\cref{prop:fixed-point-preservation}.

Numerical experiments confirm discrete consistency, second-order temporal
self-convergence of slowly evolving periodic velocity fields, and sharp
reductions in inner sweep counts across 2D/3D periodic flows and the enclosed
cavity. Parameter and timing studies link off-unit relaxation behavior to
surviving nonconserved modes and quantify how FFTs, sparse triangular solves,
and adaptive refactorizations shape net runtime gains, pointing toward
boundary-consistent iterative kinetic solvers for 3D wall-bounded flows with
reduced reliance on sparse direct factorization.

\section*{Acknowledgments}
Computational resources were provided by the High Performance Computing Platform
of Xiangtan University. The data, source code, and reproducibility instructions supporting the
findings of this study are provided in the supplementary code and data archive
accompanying this article. During the preparation of this manuscript, the
authors used Claude (Anthropic) and Codex (OpenAI) to assist with drafting,
language editing, literature retrieval, and consistency checking; after using
these tools, the authors reviewed and edited all content as needed and take full
responsibility for the content of the publication.

\Needspace{4\baselineskip}
\bibliographystyle{siamplain}
\bibliography{references}
\end{document}


\maketitle
\suppressfloats[t]
This supplement specifies the discrete velocity conventions, benchmark
configurations, timing definitions, operator and trajectory verification tests,
parameter diagnostics, and multigrid solver details supporting the main
article. The formulation, principal theorems, and proofs appear in the main
article.

\section{Extended test configurations and reproducibility details}
\label{sec:protocol}
Reproducing the numerical diagnostics and runtime comparisons requires fixing
the discrete velocity orderings, initial fields, acceptance and error criteria,
and timed execution intervals.

\subsection{Relation to the main article}
\Cref{main:sec:dc,main:sec:correctors} of the main article
derive the residual corrections, and \cref{main:alg:unified} gives
their common iteration and acceptance procedure. The main article also
presents the primary runtime comparisons and the off-design parameter study.
This supplement records the algebraic conventions, timing protocols, and
supporting verification data behind those results.

\subsection{Discrete velocity sets and lattice quadrature identities}
\label{app:lattices}
Exact moment projection and equilibrium lifting rest on the discrete velocity
indexing and quadrature weights of the underlying lattices. On D2Q9, the nine
velocity components, weights, and opposite-direction indices are ordered as
\begin{align}
 (c_{kx})_{k=0}^{8}&=(0,1,0,-1,0,1,-1,-1,1),\nonumber\\
 (c_{ky})_{k=0}^{8}&=(0,0,1,0,-1,1,1,-1,-1),\nonumber\\
 (w_k)_{k=0}^{8}&=\left(\frac49,\frac19,\frac19,\frac19,\frac19,
 \frac1{36},\frac1{36},\frac1{36},\frac1{36}\right),\nonumber\\
 (\bar k)_{k=0}^{8}&=(0,3,4,1,2,7,8,5,6).
 \label{eq:d2q9}
\end{align}
Here $c_{kx}$ and $c_{ky}$ are the Cartesian components of $\bm c_k$, $w_k$ is
its quadrature weight, and $\bar k$ indexes the opposite link
$\bm c_{\bar k}=-\bm c_k$. On D3Q19, the nineteen directions follow the
corresponding three-dimensional ordering
\begin{align}
 (c_{kx})_{k=0}^{18}&=(0,1,-1,0,0,0,0,1,-1,1,-1,1,-1,1,-1,0,0,0,0),
 \nonumber\\
 (c_{ky})_{k=0}^{18}&=(0,0,0,1,-1,0,0,1,-1,-1,1,0,0,0,0,1,-1,1,-1),
 \nonumber\\
 (c_{kz})_{k=0}^{18}&=(0,0,0,0,0,1,-1,0,0,0,0,1,-1,-1,1,1,-1,-1,1),
 \nonumber\\
 (w_k)_{k=0}^{18}&=\left(\frac13,
 \underbrace{\frac1{18},\ldots,\frac1{18}}_{6},
 \underbrace{\frac1{36},\ldots,\frac1{36}}_{12}\right),\nonumber\\
 (\bar k)_{k=0}^{18}&=(0,2,1,4,3,6,5,8,7,10,9,12,11,14,13,16,15,18,17).
 \label{eq:d3q19}
\end{align}
Here $c_{kz}$ supplies the third coordinate velocity component. Fixing
\eqref{eq:d2q9} uniquely labels the D2Q9 wall bounce-back assignments in
\eqref{main:eq:bounceback-rows}, while any consistent index permutation
leaves the discrete dynamics invariant.

Both velocity sets satisfy isotropic discrete quadrature identities through
fourth order at $\cs^2=1/3$ for coordinate indices
$\alpha,\beta,\gamma,\delta\in\{1,\ldots,d\}$:
\begin{align}
 \sum_{k=0}^{Q-1} w_k &= 1,\qquad
 \sum_{k=0}^{Q-1} w_k c_{k\alpha} = 0,\qquad
 \sum_{k=0}^{Q-1} w_k c_{k\alpha}c_{k\beta} = \cs^2\delta_{\alpha\beta},
 \label{eq:quad-012}\\
 \sum_{k=0}^{Q-1} w_k c_{k\alpha}c_{k\beta}c_{k\gamma} &= 0,\qquad
 \sum_{k=0}^{Q-1} w_k c_{k\alpha}c_{k\beta}c_{k\gamma}c_{k\delta}
 = \cs^4\bigl(\delta_{\alpha\beta}\delta_{\gamma\delta}
  +\delta_{\alpha\gamma}\delta_{\beta\delta}
  +\delta_{\alpha\delta}\delta_{\beta\gamma}\bigr),
 \label{eq:quad-34}
\end{align}
where $\delta_{\alpha\beta}$ denotes the Kronecker delta and the diagonal
fourth-moment identity holds because $c_{k\alpha}\in\{-1,0,1\}$ satisfies
$c_{k\alpha}^4=c_{k\alpha}^2$ while $3\cs^4-\cs^2=0$ at $\cs^2=1/3$.
\Cref{eq:quad-012,eq:quad-34} immediately imply
$PE=I_{d+1}$ for the projection and lift in \eqref{main:eq:linear-lift},
as well as $P\fvec^{\mathrm{eq}}=\bm m$ and $P\bm\Phi=\bm s$ behind the exact
full-source moment injection in \eqref{main:eq:collision-operating}.

\subsection{Test problems, acceptance criteria, and timing definitions}
With the moment algebra fixed, nominal benchmarks use lattice units
($\Delta x=\Delta t_{\mathrm{ac}}=1$, $\nu=1/6$, $\Lambda=1/4$, so
$\omega^+=\omega^-=1$). Setting $\Re=UN/\nu$ and $\Ma=U/\cs$ from directional
node count $N$ and characteristic speed $U$ (initial TGV amplitude or cavity
lid speed $u_{\mathrm{lid}}$) fixes each operating point.

On the two-dimensional $N\times N$ periodic lattice with fundamental
wavenumber $k_0=2\pi/N$ and node coordinates $x,y\in\{0,\ldots,N-1\}$, the
analytic incompressible Taylor--Green vortex field evolves as
\begin{align}
 u_x(x,y,t)&=-U\cos(k_0x)\sin(k_0y)\exp(-2\nu k_0^2t), \nonumber\\
 u_y(x,y,t)&=\phantom{-}U\sin(k_0x)\cos(k_0y)\exp(-2\nu k_0^2t),          \label{eq:tgv2d}\\
 p'(x,y,t)&=-\frac{\rho_0U^2}{4}
 [\cos(2k_0x)+\cos(2k_0y)]\exp(-4\nu k_0^2t), \nonumber
\end{align}
where $U=\Re\nu/N$, $p'$ is the hydrodynamic pressure fluctuation, and
$\rho_0=1$ is reference density. Populations start at $t=0$ from
$f_k^{\mathrm{eq}}(\hat p(\bm x,0),\uvec(\bm x,0))$ with
\begin{equation}
 \hat p(x,y,0)=\frac{p'(x,y,0)}{\cs^2}
 =-\frac{\rho_0U^2}{4\cs^2}
 [\cos(2k_0x)+\cos(2k_0y)].
 \label{eq:tgv2d-pressure-init}
\end{equation}
Here $\hat p=p'/\cs^2$ matches the initial pressure gradient to the solenoidal
velocity field so acoustic startup oscillations are suppressed. At
$t=T_{\mathrm{end}}$, the absolute endpoint error $e_{\mathrm{abs}}$ is the RMS
norm, over all grid points, of the velocity difference from
\eqref{eq:tgv2d}. The relative error divides $e_{\mathrm{abs}}$ by the RMS
norm of the numerical endpoint velocity, with a $10^{-12}$ denominator floor.

On the three-dimensional $N^3$ periodic D3Q19 lattice, runs start from zero
pressure perturbation ($\hat p(\bm x,0)=0$) and the solenoidal velocity field
\begin{equation}
 \begin{aligned}
 \uvec(x,y,z,0)
 &=U\bigl(\sin(k_0x)\cos(k_0y)\cos(k_0z),\\
 &\hspace{13mm}-\cos(k_0x)\sin(k_0y)\cos(k_0z),\;0\bigr),\\
 U&=\frac{\Re\nu}{N}.
 \end{aligned}
 \label{eq:tgv3d-ic}
\end{equation}
Here $z\in\{0,\ldots,N-1\}$ indexes the third periodic coordinate. The $1\%$
relative-error criterion compares the endpoint velocity with the same-grid
explicit LBM solution and normalizes their RMS difference by the initial
RMS speed $U/2$. The matched-accuracy criterion instead compares the absolute
RMS departures of the explicit solution and the solution obtained with dual-time
stepping from the prescribed
modal decay field $\uvec_{\mathrm{mod}}(\bm x,t)=\uvec(\bm x,0)\exp(-4\nu k_0^2t)$
(used as a diagnostic reference field rather than an exact three-dimensional
Navier--Stokes solution).

In the wall-bounded benchmark, the two-dimensional cavity starts impulsively
from rest $(\hat p,\uvec)=(0,\bm0)$ with lid speed $u_{\mathrm{lid}}=\Re\nu/N$,
wall separation $(N-1)\Delta x$, and ordered on-grid bounce-back closure
\eqref{main:eq:bounceback-rows}, measuring the endpoint velocity error on the
same grid against the explicit LBM reference $\uvec_{\mathrm{ref}}$ via
\begin{equation}
 E_u=\frac{1}{u_{\mathrm{lid}}}
 \left[\frac{1}{N_\Omega}\sum_{\bm x}
 \left|\uvec(\bm x)-\uvec_{\mathrm{ref}}(\bm x)\right|^2\right]^{1/2}.
 \label{eq:cavity-error}
\end{equation}
Here $N_\Omega=N^2$ counts all lattice nodes in the cavity and $E_u$ defines
the normalized endpoint velocity error tested against the $1\%$ and $0.1\%$
criteria. \Cref{tab:protocol} lists the grids, integration intervals,
physical-step counts, and residual tolerances used in the parameter tests, with
$\dt=T_{\mathrm{end}}/S$ and $\operatorname{nint}(x)=\lfloor x+1/2\rfloor$ for
$x>0$.

\begin{table}[!tbp]
\centering
\caption{Test configurations and candidate parameter sets. The physical step is
$\dt=T_{\mathrm{end}}/S$, with additional boundary tests at the extended $S$
values in parentheses. \Cref{sec:full-results,sec:parameter-range} list
fixed-step and parameter-scan configurations.}
\label{tab:protocol}
\footnotesize
\renewcommand{\arraystretch}{0.96}
\begin{tabular}{@{}>{\raggedright\arraybackslash}p{0.10\textwidth}
                    >{\raggedright\arraybackslash}p{0.48\textwidth}
                    >{\raggedright\arraybackslash}p{0.30\textwidth}@{}}
\toprule
case & grids, Reynolds numbers, and horizon & candidate step and tolerance sets\\
\midrule
2D TGV &
$\Re=1$: $N=32$--$1024$ by factors of two;
$\Re=100$: $N=128$--$1024$;
$N=256$: $\Re=10^{-4}$--$10^2$ by decades.
$T_{\mathrm{end}}=\operatorname{nint}(0.15N^2)$. &
$S=8,16,32,64,128$ (extended to $4,256$).
$\varepsilon=10^{-6},10^{-8},10^{-10}$;
$K_{\max}=300$.\\
3D TGV &
$\Re=1$: $N=32,64,128$;
$\Re=100$: $N=64,128$.
$T_{\mathrm{end}}=\operatorname{nint}(0.075N^2)$. &
$S=4,5,8,16,32$ (extended to $3,64$).
$\varepsilon=10^{-6},10^{-8},10^{-10},10^{-12}$;
$K_{\max}=300$.\\
cavity &
$N=64,128$ at $\Re=10,100$ for timing;
$N=32,256$ for single-step mechanism tests.
$T_{\mathrm{end}}=\operatorname{nint}(1.5N^2)$. &
$S=4,5,8,16,32$ (extended to $3,64$).
$\varepsilon=10^{-6},10^{-8},10^{-10},10^{-12}$;
$K_{\max}=4000$.\\
\bottomrule
\end{tabular}
\end{table}

Physical-step acceptance tests the population residual
\eqref{main:eq:residual}, requiring $R_\ell<\varepsilon$ after at least
$n_{\min}=2$ sweeps in 2D periodic runs and $n_{\min}=1$ in 3D and cavity runs,
while rejecting the trajectory if any step fails this criterion within
$K_{\max}$ sweeps. The matched-accuracy criterion requires
$e_{\mathrm{abs}}\leq 3e_{\mathrm{explicit}}$. The relative-error criteria require
the endpoint relative error (or $E_u$ in the cavity) to be below $1\%$ or
$0.1\%$. Stopping and startup conventions for the reference solvers are
summarized in \cref{sec:comparison-protocols}.

For a sequence of nonlinear inner iterations ending at sweep
$\ell_{\mathrm f}$ without a Jacobian refresh, the observed residual
contraction is measured by the geometric mean of successive residual ratios,
\begin{equation}
 \rho_{\mathrm{obs}}
 =\exp\!\left[
 \frac{1}{m_\rho}\sum_{j=\ell_{\mathrm f}-m_\rho+1}^{\ell_{\mathrm f}}
 \log\!\left(\frac{R_j}{R_{j-1}}\right)\right].
 \label{eq:rho-observed}
\end{equation}
For $R_j>0$ denoting the population residual RMS after sweep $j$, the window
length is $m_\rho=\min(10,\ell_{\mathrm f}-1)$ in two-dimensional tests
and $m_\rho=\min(8,\ell_{\mathrm f}-1)$ in three-dimensional tests,
evaluated over refresh-free segments with $\ell_{\mathrm f}\ge3$ and restarted
after any adaptive Jacobian refresh.

Using these diagnostics, 2D periodic tables compare the
macroscopic-shadow-corrected lattice Boltzmann method (MSC-LBM;
\cref{main:sec:shadow}) against two
multigrid solvers targeting the fine-grid TRT/BDF problem: FAS W-cycle
multigrid smoothed by Picard iteration (MG) and weighted-Jacobi FAS multigrid
(Gsell-type)~\cite{Gsell2020} (\cref{app:comparison-mg}). Three-dimensional
tests compare MSC-LBM with unaccelerated Picard iteration. Cavity tests compare
Picard iteration with the kinetic solvers that reuse or adaptively update the
reference Jacobian in \cref{main:sec:walls}. In
\cref{main:tab:tgv2d,main:tab:tgv3d,main:tab:cavity}, each reported
$(S,\varepsilon)$ pair minimizes $W$ over the tested configurations satisfying
the stated error criterion; the weights defining $W$ are given in
\cref{sec:parameter-range}.

Timing runs execute as single-core serial processes pinned to one physical core
of an exclusive dual-socket Intel Xeon Platinum 8358 node ($2\times32$ cores,
hyper-threading disabled), recording the minimum elapsed time over two explicit
runs and four runs with dual-time stepping (four repetitions for both the
explicit baseline and the tested solvers in \cref{tab:total-runtime}). The Fortran solvers
are compiled with GNU Fortran~13.2.0 using \texttt{-O3},
\texttt{-march=x86-64-v3}, \texttt{-mtune=generic}, \texttt{-funroll-loops},
and \texttt{-ffree-\allowbreak line-\allowbreak length-\allowbreak none} with
fast-math disabled, linking sequential LP64 Intel oneAPI MKL PARDISO for sparse
cavity LU solves and a self-contained radix-2 FFT for periodic transforms.

Timed intervals match the two operational ratios reported in the main article.
For periodic flows, timing encloses the entire physical time-stepping loop
(kinetic sweeps, moment projections, forward and inverse FFTs with bit-reversal
and twiddle evaluation, Schur solves, and equilibrium lifting, excluding
one-time allocation and terminal error evaluation) to define
\begin{equation}
 \Swall=\frac{t_{\mathrm{explicit}}}{t_{\mathrm{scheme}}}.
 \label{eq:swall}
\end{equation}
Here $t_{\mathrm{explicit}}$ and $t_{\mathrm{scheme}}$ denote the minimum
serial loop runtimes of the explicit baseline and the solver with dual-time
stepping. For the cavity, timing isolates the physical
marching phase via
\begin{equation}
 \Smarch=
 \frac{t_{\mathrm{explicit,march}}}{t_{\mathrm{scheme,march}}}.
 \label{eq:smarch}
\end{equation}
In \eqref{eq:smarch}, $t_{\mathrm{explicit,march}}$ times the explicit marching
loop, while $t_{\mathrm{scheme,march}}$ includes marching kinetic sweeps,
sparse PARDISO triangular back-solves, and adaptive Jacobian reassemblies and
LU refactorizations during marching. Prior to step~1, the shared cavity kinetic
driver executes a preliminary two-step Picard pass (used only when preparing a
nonzero reference state), assembles and factors $I-L$ at the rest state
$(\hat p,\uvec)=(1,\bm0)$, and resets the populations and BDF histories to
rest; this pre-march initialization is excluded from $\Smarch$ and included in
total solver runtime, defining the total solver time ratio across four
repetitions of both solvers as
\begin{equation}
 \Stotal=
 \frac{t_{\mathrm{explicit,total}}}{t_{\mathrm{scheme,total}}}.
 \label{eq:stotal}
\end{equation}

\subsection{Comparison protocols and implementation conventions}\label{sec:comparison-protocols}
The numerical study combines two complementary timing protocols.
\Cref{main:tab:tgv2d,main:tab:tgv3d,main:tab:cavity} report work-selected
marching ratios ($\Swall$ and $\Smarch$) across the candidate $(S,\varepsilon)$
sets in \cref{tab:protocol}, including configurations where no candidate meets
the accuracy criterion or where auxiliary overhead yields ratios below unity.
In these comparisons, \cref{main:alg:unified} evaluates the population residual
\eqref{main:eq:residual} at the sweep input $\fvec$ prior to accepting
$\fvec_T=T(\fvec)$, whereas the FAS multigrid solvers evaluate the fine-grid
defect at the cycle output (\cref{app:comparison-mg}), and two-dimensional
parameter-search runs include pre-march pressure relaxation or linear step
prediction depending on the solver branch. Complementing these parameter-search
tables, \cref{tab:total-runtime} fixes $(S,\varepsilon)$, startup
initialization, and quadratic step prediction across matched exclusive-node
repetitions to report both marching ratios and total solver time ratios
$\Stotal$ including setup and initial sparse factorization.

\section{Operator and trajectory verification}
\label{sec:full-results}
This section examines numerical agreement between solvers, analytical Jacobian
accuracy, and temporal self-convergence on fixed spatial grids.

\subsection{Discrete consistency and temporal accuracy}
\label{sec:verification}
Under fixed history and boundary data with $a,\nu,\cs^2>0$ and nonsingular
wall matrices $I-L$, \cref{main:prop:fixed-point-preservation} shows
that residual correction leaves the discrete BDF fixed-point set unchanged. On
a single interior BDF2 step of 2D TGV decay
initialized from analytic histories at $t=0$ and $\dt$ ($N=128$, $\Re=100$,
$\dt=38.9$, $t=2\dt=77.8$, $\varepsilon=10^{-12}$, $K_{\max}=600$), MSC-LBM and
Picard velocity fields agree to $3.57\times10^{-11}$ in absolute vector RMS
and $4.15\times10^{-10}$ in relative RMS. In two-step (BDF1 startup plus BDF2)
runs at $\Re=1$ comparing MSC-LBM and Picard at
$\varepsilon=10^{-12}$ ($K_{\max}=20000$) and $\varepsilon=10^{-14}$
($K_{\max}=40000$), every step converges below $\varepsilon$: in 2D
($N=128,\dt=100,T_{\mathrm{end}}=200$), the full-population RMS difference
drops from $5.27\times10^{-11}$ to $5.27\times10^{-13}$ (velocity vector RMS
from $3.19\times10^{-10}$ to $3.19\times10^{-12}$); in 3D
($N=32,\dt=20,T_{\mathrm{end}}=40$), it drops from $4.61\times10^{-12}$ to
$4.25\times10^{-14}$ (velocity vector RMS from $6.54\times10^{-11}$ to
$6.06\times10^{-13}$). Both iterations
test the same population residual~\eqref{main:eq:residual}, and the shadow
increment vanishes at $\bm r=\bm0$.

Operator-level verification confirms both collision-branch equivalence and
wall-Jacobian exactness. At $\omega^+=\omega^-=1$, where
\eqref{main:eq:collision-operating} coincides algebraically with unit-$\tau$
BGK collision, the TRT and BGK code paths differ by at most
$1.0842021725\times10^{-18}$ after the BDF1 startup step and
$1.8555027308\times10^{-13}$ after 100 physical steps (one BDF1 step followed
by 99 BDF2 steps). In the $N=64$, $\Re=10$,
$\dt=38.9$ cavity, forward-difference evaluation of the complete sweep
$T_{q,\bm h_q,\Gamma}$, including the ordered bounce-back rules in
\eqref{main:eq:bounceback-rows}, matches the analytical global Jacobian $L$
to $1.234\times10^{-6}$ at step $\delta=10^{-6}$. Because reconstruction is
affine in $\fvec$, equilibrium and source distributions are quadratic in
$\bm m$, and streaming plus ordered bounce-back assignments act as affine row
copies, all third derivatives of $T$ vanish at unit relaxation, making the
exact-arithmetic Taylor remainder for $(D_\delta T)_{pq}=[T_p(\bar\fvec+\delta\bm e_q)-T_p(\bar\fvec)]/\delta$
purely first order in $\delta$ on D2Q9 ($\cs^2=1/3$):
\begin{equation}
 (D_\delta T)_{pq}-L_{pq}
 =\frac{\delta}{2}\frac{\partial^2T_p}{\partial f_q^2},\qquad
 \max_{p,q}\left|(D_\delta T)_{pq}-L_{pq}\right|
 =\frac{4}{3}\frac{|1-a_2|}{(1+a_2/2)^2}\,\delta .
 \label{eq:fd-truncation-supp}
\end{equation}
Here $p$ and $q$ index global population entries, $a_2=3/(2\dt)$ is the BDF2
coefficient, and the maximum is attained at rest weight $w_0=4/9$ along a
diagonal velocity perturbation $|\bm c_q|^2=2$ while inhomogeneous lid shifts
differentiate to zero. Evaluating \eqref{eq:fd-truncation-supp} at
$a_2=3/77.8$ and $\delta=10^{-6}$ predicts $1.2339\times10^{-6}$, matching the
measured discrepancy to four significant digits. In the sparse
cavity solver, directional forward differences across BDF1 rest, BDF2 rest,
and BDF2 inhomogeneous two-level-history states at
$\delta\in\{10^{-4},10^{-6},10^{-8}\}$ match the assembled sparse CSR matrix
$I-L$ once the exact quadratic Taylor remainder is subtracted: componentwise
compensated errors remain within $0.081$ of the outward-rounded floating-point
error bound across all $9N^2=576$ columns at $N=8$ ($5184$ checks) and within
$0.124$ across $12$ interior, wall, corner, and full-field directions at $N=64$
($108$ checks). Across full cavity trajectories, \cref{tab:verification} confirms that
Picard, frozen-kinetic, and adaptive-kinetic solves reach minimum endpoint
errors matching at all three displayed digits in $(64,10)$ and $(64,100)$, and
within about $2\%$ at $(128,100)$.

\begin{table}[!tbp]
\centering
\caption{Minimum cavity velocity errors reached in the tested
configurations. Errors use \eqref{eq:cavity-error} at the actual
terminal times.}
\label{tab:verification}
\small
\begin{tabular}{lccc}
\toprule
cavity case $(N,\Re)$ & Picard & kinetic (frozen) & kinetic (adaptive)\\
\midrule
$(64,10)$   & $3.04\times10^{-6}$ & $3.04\times10^{-6}$ & $3.04\times10^{-6}$\\
$(64,100)$  & $6.53\times10^{-6}$ & $6.53\times10^{-6}$ & $6.53\times10^{-6}$\\
$(128,100)$ & $4.82\times10^{-7}$ & $4.91\times10^{-7}$ & $4.90\times10^{-7}$\\
\bottomrule
\end{tabular}
\end{table}

Temporal self-convergence is tested by holding the spatial mesh fixed and
halving the physical time step. The component-wise RMS difference between the
resulting terminal velocity fields and the observed order are
\begin{equation}
 D_i=\left[\frac{1}{dN_\Omega}\sum_{\bm x}
 \left|\uvec_{\dt_i}(\bm x)-\uvec_{\dt_i/2}(\bm x)\right|^2\right]^{1/2},
 \qquad p_i=\log_2\!\left(\frac{D_i}{D_{i+1}}\right).
 \label{eq:observed-order}
\end{equation}
Here $d\in\{2,3\}$ is the spatial dimension, $\uvec_{\dt_i}$ denotes the
terminal velocity field advanced with step $\dt_i$, and $p_i$ is the observed
Richardson order reported in \cref{main:tab:temporal} using MSC-LBM at
$\varepsilon=10^{-12}$ and $K_{\max}=20000$. Every physical step across all
eight runs satisfies $R_\ell<10^{-12}$ (all-step maximum $9.99\times10^{-13}$),
with terminal-step residuals lying between $3.11\times10^{-14}$ and
$9.57\times10^{-13}$ in 2D ($N=128,\Re=1,T_{\mathrm{end}}=1600$,
$\dt\in\{200,100,50,25\}$, initialized from \eqref{eq:tgv2d-pressure-init}) and between
$3.65\times10^{-13}$ and $9.31\times10^{-13}$ in 3D
($N=32,\Re=1,T_{\mathrm{end}}=160$, $\dt\in\{16,8,4,2\}$). The observed orders
($2.178,2.062$ in 2D; $2.111,2.069$ in 3D) verify second-order temporal
self-convergence for smooth, slowly evolving velocity fields on periodic
domains.

\subsection{Summary of runtime tables and diagnostic cross-references}
The preceding checks verify discrete consistency and temporal accuracy. Using
the definitions in \cref{sec:protocol}, the main article reports the
work-selected two-dimensional periodic, three-dimensional periodic, and
enclosed-cavity runtime comparisons in \cref{main:tab:tgv2d,main:tab:tgv3d,main:tab:cavity}, while
\cref{sec:parameter-range} provides the off-design contraction
measurements, fixed-step total solver runtimes (\cref{tab:total-runtime}),
and large-step residual-tolerance diagnostics
(\cref{tab:residual-tolerance}).

\section{Additional parameter diagnostics and cost definitions}
\label{sec:parameter-range}
At unit TRT relaxation with $\nu=1/6$, the relation
$\Ma=\sqrt{3}\Re/(6N)$ defines the algebraic Mach contour
$\Ma=0.34\iff\Re/N=1.1778\approx1.2$ across the $(N,\Re)$ plane.

Complementing \cref{main:sec:range-main}, this section summarizes the
sampled periodic contraction measurements and defines the weighted work proxy
$W$.

On smooth periodic TGV fields at unit TRT relaxation ($\nu=1/6$,
$\Lambda=1/4$), $\Ma=\Re\nu/(N\cs)$ decreases as $N^{-1}$ under fixed-$\Re$
grid refinement, reducing $\rho_{\mathrm{obs}}$ down to $0.042$ at $N=1024$
while $O(N^{d+2})$ explicit work growth drives the rising $\Swall$ trends at
$\Re=1$; on general even periodic meshes, the uncorrected axial Nyquist mode
retains eigenvalue $-\gamma$ ($|\gamma|\to1$ as $a\to0^+$), precluding a
step-size-independent all-mode contraction floor. Away from unit relaxation on
the interior BDF2 step ($N=128$, $\dt=38.9$, $\Ma=0.1$,
$\varepsilon=10^{-10}$), varying $\Lambda\in\{1/4,1/6,1/2,1/12,1\}$ at
$\nu=1/6$ yields $\rho_{\mathrm{obs}}=0.109,0.130,0.276,0.497,0.596$, tracking
the odd-sector retention $|1-\omega^-|=0,0.200,0.333,0.500,0.600$. Varying
$\nu\in\{0.13,0.11,0.10,0.095\}$ along $\Lambda=1/4$ activates both parity
sectors with $\eta(\nu)=0.124,0.205,0.250,0.274$ and produces
$\rho_{\mathrm{obs}}=0.266,0.410,0.530,0.716$ before divergence at $\nu=0.09$
($\eta=0.299$): because streaming couples surviving even nonconserved modes
back into the macroscopic correction, $\rho_{\mathrm{obs}}$ exceeds the
single-collision retention factor $\eta(\nu)$ itself rather than obeying a
multiplicative $\eta(\nu)$ bound.

Configuration selection uses the prescribed work proxy $W$. The associated
work ratio compares the explicit step count with this proxy,
\begin{equation}
 \Swork=\frac{n_{\mathrm{ref}}}{W},\qquad
 n_{\mathrm{ref}}=\frac{T_{\mathrm{end}}}{\Delta t_{\mathrm{ac}}}.
 \label{eq:swork}
\end{equation}
Here $n_{\mathrm{ref}}$ is the explicit acoustic step count and $W$ is the
prescribed work proxy: in periodic runs, $W$ assigns unit weight to each
fine-grid $Q$-population sweep, weight $4^{-\ell}$ to each level-$\ell$
coarse-grid sweep in the 2D multigrid hierarchy, and unit weight to each
periodic spectral correction (which executes $2(d+1)$ complex FFTs, namely six
$N^2$ transforms in 2D and eight $N^3$ transforms in 3D, plus retained-mode
Schur solves); in cavity runs, $W$ counts fine-grid kinetic sweeps alone.
By contrast, $\Swall$ in \eqref{eq:swall} measures actual periodic loop
time including FFTs, and $\Smarch$ in \eqref{eq:smarch} includes PARDISO
triangular solves with the LU factors of the $9N^2$-unknown system, including
fill-in, and adaptive LU refactorizations during physical time advancement.
These additional operations explain the difference between $\Swork$ and
$\Smarch$ discussed in \cref{main:sec:cavity}. The initial
rest-state factorization is excluded from this timing interval, as defined in
\cref{main:sec:protocol}.

\subsection{Fixed-step total solver runtimes and cavity setup breakdown}
\label{sec:total-runtime}
To quantify solver initialization, the preliminary two-step Picard pass, matrix
assembly, and initial sparse LU factorization alongside physical marching,
\cref{tab:total-runtime} reports four-repetition minimum marching and total
solver runtimes under fixed $(S,\varepsilon)$ protocols. Each row pairs the
solver using dual-time stepping with an explicit baseline measured over four
repetitions on the same pinned physical core of the same exclusive node; the
maximum four-repetition spread in total solver time across all listed
configurations is $1.10\%$.

\begin{table}[!tbp]
\centering
\caption{Four-repetition minimum marching and total solver runtimes (listed as
\emph{march / total}, in seconds) and resulting march and total time ratios
($\Stotal$, \eqref{eq:stotal}) under fixed $(S,\varepsilon)$ protocols.
Two-dimensional periodic runs start directly from \eqref{eq:tgv2d-pressure-init}
with quadratic step prediction.}
\label{tab:total-runtime}
\footnotesize
\setlength{\tabcolsep}{4.5pt}
\renewcommand{\arraystretch}{0.95}
\begin{tabular}{llrrrr}
\toprule
 & & \multicolumn{2}{c}{march / total time (s)} & & \\
\cmidrule(lr){3-4}
configuration & solver & explicit & tested solver & march ratio & $\Stotal$\\
\midrule
\multicolumn{6}{l}{\emph{2D periodic TGV, matched-accuracy criterion ($\Re=1$, $\varepsilon=10^{-6}$)}}\\
$N=128$, $S=16$ & MSC-LBM & 1.273 / 1.278 & 0.1544 / 0.1618 & 8.24  & 7.89\\
$N=256$, $S=32$ & MSC-LBM & 21.36 / 21.37 & 1.315 / 1.342   & 16.24 & 15.93\\
$N=512$, $S=64$ & MSC-LBM & 346.5 / 346.6 & 11.63 / 11.73   & 29.80 & 29.55\\
\midrule
\multicolumn{6}{l}{\emph{2D periodic TGV, $1\%$ relative-error criterion ($S=8$, $\varepsilon=10^{-6}$)}}\\
$(128,1)$   & MSC-LBM & 1.273 / 1.277 & 0.0844 / 0.0918 & 15.09  & 13.91\\
$(256,0.1)$ & MSC-LBM & 21.33 / 21.35 & 0.3284 / 0.3545 & 64.96  & 60.23\\
$(256,1)$   & MSC-LBM & 21.33 / 21.35 & 0.3278 / 0.3544 & 65.09  & 60.25\\
$(256,100)$ & MSC-LBM & 21.35 / 21.37 & 0.5421 / 0.5679 & 39.38  & 37.63\\
$(512,1)$   & MSC-LBM & 346.8 / 346.9 & 1.457 / 1.557   & 238.06 & 222.81\\
\midrule
\multicolumn{6}{l}{\emph{3D periodic TGV, $1\%$ relative-error criterion ($\Re=1$, $S=16$, $\varepsilon=10^{-10}$)}}\\
$N=64$  & Picard  & 14.42 / 14.45 & 87.81 / 87.85 & 0.16 & 0.16\\
$N=64$  & MSC-LBM & 14.42 / 14.45 & 16.28 / 16.32 & 0.89 & 0.89\\
$N=128$ & MSC-LBM & 464.2 / 464.5 & 103.9 / 104.3 & 4.47 & 4.45\\
\midrule
\multicolumn{6}{l}{\emph{2D enclosed cavity, $1\%$ relative-error criterion ($\varepsilon=10^{-10}$)}}\\
$(64,10)$, $S=16$   & Picard           & 0.877 / 0.877 & 5.35 / 5.35   & 0.16 & 0.16\\
$(64,10)$, $S=16$   & kinetic frozen   & 0.877 / 0.877 & 0.392 / 2.35  & 2.24 & 0.37\\
$(64,10)$, $S=16$   & kinetic adaptive & 0.877 / 0.877 & 0.393 / 2.35  & 2.23 & 0.37\\
$(128,10)$, $S=64$  & Picard           & 14.97 / 14.97 & 45.07 / 45.07 & 0.33 & 0.33\\
$(128,10)$, $S=64$  & kinetic frozen   & 14.97 / 14.97 & 3.663 / 12.88 & 4.09 & 1.16\\
$(128,10)$, $S=64$  & kinetic adaptive & 14.97 / 14.97 & 3.664 / 12.89 & 4.09 & 1.16\\
$(128,100)$, $S=64$ & Picard           & 14.93 / 14.93 & 63.82 / 63.82 & 0.23 & 0.23\\
$(128,100)$, $S=64$ & kinetic frozen   & 14.93 / 14.93 & 24.64 / 35.50 & 0.61 & 0.42\\
$(128,100)$, $S=64$ & kinetic adaptive & 14.93 / 14.93 & 15.86 / 26.72 & 0.94 & 0.56\\
\bottomrule
\end{tabular}
\end{table}

In the $(128,10)$ cavity run at $S=64$ ($\dt=384$), the representative
minimum-total repetition of the frozen kinetic solver ($12.88\,\mathrm{s}$
total) spends $9.22\,\mathrm{s}$ ($71.5\%$) in pre-march initialization---comprising
$7.22\,\mathrm{s}$ ($56.0\%$) in the preliminary two-step Picard pass,
$1.97\,\mathrm{s}$ ($15.3\%$) in the initial PARDISO sparse LU factorization,
and $0.030\,\mathrm{s}$ ($0.2\%$) in CSR matrix assembly---followed by
$3.67\,\mathrm{s}$ ($28.5\%$) across the $64$ physical steps ($2.4$ mean sweeps
per step versus $503.7$ for Picard). At $(128,100)$ and $S=64$, one adaptive
refresh reduces total solver runtime from $35.50\,\mathrm{s}$ (frozen) and
$63.82\,\mathrm{s}$ (Picard) to $26.72\,\mathrm{s}$ ($2.39$ times faster than
Picard in total time, though still above the $14.93\,\mathrm{s}$ explicit
run). In the two-dimensional periodic matched-accuracy series ($N=128,256,512$
with $S=16,32,64$), the MSC-LBM to explicit endpoint velocity RMS error ratios
are $0.81,0.82,1.19$, consistent with resolving the slow hydrodynamic velocity
evolution on time scales $\dt\gg\Delta t_{\mathrm{ac}}$.

\subsection{Residual-tolerance scaling and large-step termination}
\label{sec:residual-scaling}
\Cref{main:prop:moment-bottleneck} and \eqref{main:eq:zero-mode-scale} show
that raw conserved-moment residuals are attenuated by $O(a)=O(\dt^{-1})$
relative to macroscopic state errors at large physical steps. On
two-dimensional periodic TGV decay at $\Re=1$ starting directly from
\eqref{eq:tgv2d-pressure-init} with $(S,\varepsilon)=(8,10^{-6})$ and
$K_{\max}=300$ ($T_{\mathrm{end}}=2458,\dt=307.25$ at $N=128$;
$T_{\mathrm{end}}=9830,\dt=1228.75$ at $N=256$), unaccelerated Picard satisfies
$R_\ell<10^{-6}$ in $1$--$2$ sweeps per step ($11$ total sweeps across the eight
steps) yet retains endpoint relative velocity errors of $84.3\%$ at $N=128$ and
$85.7\%$ at $N=256$ ($537.4\%$ and $598.5\%$ when normalized by the analytic
endpoint velocity RMS, with fundamental-mode amplitude ratios $6.37$ and
$6.99$). Under the same stopping threshold, FAS W-cycle multigrid smoothed by
Picard iteration (MG) terminates in two cycles per step ($16$ total cycles) with
endpoint relative errors of $184.3\%$ at $N=128$ and $1293.3\%$ at $N=256$
($64.8\%$ and $92.8\%$ against the analytic endpoint RMS, with over-damped
fundamental-mode amplitude ratios $0.352$ and $0.0718$ and orthogonal shape
deviations below $0.11\%$). By inverting the low-wavenumber moment defect prior
to checking $R_\ell<\varepsilon$, MSC-LBM reaches $0.821\%$ ($N=128$) and
$0.836\%$ ($N=256$) relative error in two sweeps per step ($16$ total sweeps).

\begin{table}[!tbp]
\centering
\caption{Step-size and inner-tolerance interaction for MSC-LBM on
two-dimensional periodic TGV decay at $(N,\Re,T_{\mathrm{end}})=(256,1,9830)$
with $K_{\max}=300$, initialized directly from \eqref{eq:tgv2d-pressure-init}.
At $(S,\varepsilon)=(32,10^{-10})$, steps $24,28,31$ reach $K_{\max}$ at
$R_\ell=1.04\times10^{-10}$--$1.12\times10^{-10}$; the trajectory is rejected
by the all-step criterion and its terminal field errors are included for
comparison.}
\label{tab:residual-tolerance}
\footnotesize
\setlength{\tabcolsep}{4.5pt}
\renewcommand{\arraystretch}{0.95}
\begin{tabular}{rrrrrrl}
\toprule
$S$ & $\dt$ & $\varepsilon$ & converged steps & total sweeps & $e_{\mathrm{abs}}$ & $e_{\mathrm{rel}}$ status\\
\midrule
$8$  & $1228.75$ & $10^{-6}$  & $8/8$   & $16$   & $5.39\times10^{-7}$ & $0.836\%$ (accepted)\\
$8$  & $1228.75$ & $10^{-10}$ & $8/8$   & $25$   & $5.56\times10^{-7}$ & $0.862\%$ (accepted)\\
$32$ & $307.19$  & $10^{-6}$  & $32/32$ & $64$   & $5.18\times10^{-9}$ & $0.00811\%$ (accepted)\\
$32$ & $307.19$  & $10^{-10}$ & $29/32$ & $1932$ & $1.66\times10^{-8}$ & $0.0259\%$ (rejected)\\
$64$ & $153.59$  & $10^{-6}$  & $64/64$ & $128$  & $1.12\times10^{-8}$ & $0.0175\%$ (accepted)\\
$64$ & $153.59$  & $10^{-10}$ & $64/64$ & $133$  & $1.50\times10^{-9}$ & $0.00235\%$ (accepted)\\
\bottomrule
\end{tabular}
\end{table}

\Cref{tab:residual-tolerance} further shows how inner tolerance interacts
with outer BDF truncation error for MSC-LBM at $(N,\Re)=(256,1)$: at
$\varepsilon=10^{-6}$, refining from $S=32$ to $S=64$ slightly increases
endpoint error ($0.00811\%$ to $0.0175\%$) because algebraic stopping error at
$R_\ell\le5.33\times10^{-8}$ exceeds temporal discretization error, whereas
tightening $\varepsilon$ to $10^{-10}$ restores monotone error reduction at
$S=64$ ($0.00235\%$).

\FloatBarrier

\section{Multigrid comparison details}
\label{app:comparison-mg}
To specify the coarse-grid sweep weights in $W$ and document the 2D periodic
reference solvers introduced in \cref{main:sec:protocol}, we record the
nonlinear full-approximation-scheme (FAS) hierarchy targeting the fine-grid
TRT/BDF equation. Indexing the fine lattice as $g=0$, each coarser level $g\ge1$
scales resolution, physical step, and lattice viscosity simultaneously as
\[
 N_g=N/2^g,\qquad
 \dt_g=\dt/2^g,\qquad
 \nu_g=\nu/2^g .
\]
Here $N_g$, $\dt_g$, and $\nu_g$ define the level-$g$ grid size, BDF step, and
kinematic viscosity, fixing the coarse relaxation times by design to
$\tau_g^+=1/2+3\nu_g$ and $\tau_g^-=1$ (so $\Lambda_g=3\nu_g/2$ varies on
$g\ge1$).

Populations, BDF moment histories, and residuals transfer to coarser levels
via recursive $2\times2$ cell-average restriction $\mathcal I_g^{g+1}$.
Writing the level-$g$ fixed-history sweep as $T_g$, setting
$F_g(\fvec)=\fvec-T_g(\fvec)$, and initializing the fine-level forcing to
$\bm d_0=\bm0$, restriction assembles the coarse problem
\begin{align}
 \bm q_g&=F_g(\fvec_g)-\bm d_g, \nonumber\\
 \fvec_{g+1}^{(0)}
  &=\mathcal I_g^{g+1}\fvec_g, \nonumber\\
 \bm d_{g+1}
  &=F_{g+1}\!\left(\fvec_{g+1}^{(0)}\right)
    -\mathcal I_g^{g+1}\bm q_g .
 \label{eq:fas-transfer}
\end{align}
With level-$g$ FAS defect $\bm q_g$ (satisfying $\bm q_0=-\bm r$ on the fine
grid), restricted coarse state $\fvec_{g+1}^{(0)}$, and coarse FAS forcing
$\bm d_{g+1}$, vanishing fine-grid defect ($\bm q_0=\bm0$) makes
$\fvec_{g+1}^{(0)}$ an exact fixed point of the forced coarse problem.
The fine-grid cycle makes one call to the first coarse level. Within the
coarse hierarchy, each noncoarsest level makes two recursive calls to the
next coarser level, down to a $16^2$ coarsest lattice. On return,
cell-centered bilinear periodic prolongation $\mathcal P_{g+1}^{g}$
injects the coarse correction
\begin{equation}
 \fvec_g\leftarrow\fvec_g+
 \mathcal P_{g+1}^{g}
 \left(\fvec_{g+1}-\fvec_{g+1}^{(0)}\right).
 \label{eq:fas-return}
\end{equation}
Here $\fvec_{g+1}$ denotes the post-cycle coarse iterate, so subtracting
$\fvec_{g+1}^{(0)}$ prolongates only the coarse increment.

Smoothing on level $g$ applies either unrelaxed Picard iteration or damped
Jacobi relaxation,
\begin{align}
 \fvec_g&\leftarrow T_g(\fvec_g)+\bm d_g, \nonumber\\
 \fvec_g&\leftarrow
 (1-\lambda)\fvec_g+\lambda[T_g(\fvec_g)+\bm d_g],
 \qquad \lambda=0.6.
 \label{eq:fas-smoothers}
\end{align}
Here $\lambda=0.6$ is the under-relaxation factor used by the Gsell-type
control solver, which executes $2$ pre-sweeps and $2$ post-sweeps on each
noncoarsest level, and $4$ sweeps per visit to the coarsest grid. The W-cycle with Picard iteration as its
smoother (MG) uses $2$
fine-grid pre- and post-sweeps, $g+1$ pre- and post-sweeps on coarse level $g$,
and $2(g+1)$ sweeps per visit to the $16^2$ coarsest grid.

\bibliographystyle{siamplain}
\bibliography{references}